\documentclass{raa}            % referee version: for submission

\usepackage{graphicx,times}             %for PS/EPS graphics inclusion, new
\usepackage{natbib}
\usepackage{amssymb,amsmath}
\usepackage{lipsum}
\usepackage{mathtools}
\usepackage{cuted}
\usepackage{xcolor}
\bibpunct{(}{)}{;}{a}{}{,}

\definecolor{dkblue}{RGB}{54, 86, 169}

\usepackage[pagebackref=true]{hyperref}

\begin{document}

  \title{FAST observations of an extremely active episode of FRB 20201124A: III. Polarimetry}
%   \subtitle{I. Place Your Subtitle Here}

   \volnopage{Vol.0 (20xx) No.0, 000--000}      %%preserved for Editor. DOn't remove!
   \setcounter{page}{1}          %%starting page, preserved for Editor. DOn't remove!

   \author{
    Jin-Chen Jiang\inst{*1,2},
    Wei-Yang Wang\inst{1,3},
    Heng Xu\inst{1,2,3},
    Jiang-Wei Xu\inst{1,2,3},
    Chun-Feng Zhang\inst{1,2,3},
    Bo-Jun Wang\inst{1,2,3},
    De-Jiang Zhou\inst{2,4},
    Yong-Kun Zhang\inst{2,4},
    Jia-Rui Niu\inst{2,4},
    Ke-Jia Lee\inst{*1,2,3},
    Bing Zhang\inst{*5},
    Jin-Lin Han\inst{2,4,6},
    Di Li\inst{2,4,7},
    Wei-Wei Zhu\inst{2},
    Zi-Gao Dai\inst{8,9},
    Yi Feng\inst{7},
    Wei-Cong Jing\inst{2,4},
    Dong-Zi Li\inst{10},
    Rui Luo\inst{11},
    Chen-Chen Miao\inst{2,4},
    Chen-Hui Niu\inst{2},
    Chao-Wei Tsai\inst{2},
    Fa-Yin Wang\inst{8,9},
    Pei Wang\inst{2},
    Ren-Xin Xu\inst{1,3},
    Yuan-Pei Yang\inst{12},
    Zong-Lin Yang\inst{2,4},
    Ju-Mei Yao\inst{2}
    and Mao Yuan\inst{2,4}
   }
\institute{
Department of Astronomy, Peking University, Beijing 100871, China; \textit{jiangjinchen@pku.edu.cn}\\%1
\and
National Astronomical Observatories, Chinese Academy of Sciences, 20A Datun Road, Chaoyang District, Beijing 100101, China\\%2
\and
Kavli Institute for Astronomy and Astrophysics, Peking University, Beijing 100871, China\\%3
\and
School of Astronomy, University of Chinese Academy of Sciences, Beijing 100049, China\\%4
\and
Department of Physics and Astronomy, University of Nevada, Las Vegas, NV 89154, USA;  \textit{bing.zhang@unlv.edu}\\%5
\and
CAS Key Laboratory of FAST, NAOC, Chinese Academy of Sciences, Beijing 100101, China\\%6
\and
Research Center for Intelligent Computing Platforms, Zhejiang Laboratory, Hangzhou 311100, China\\%9
\and
School of Astronomy and Space Science, Nanjing University, Nanjing 210093, China\\%7
\and
Key Laboratory of Modern Astronomy and Astrophysics (Nanjing University), Ministry of Education, China\\%8
\and
Cahill Center for Astronomy and Astrophysics, MC 249-17 California Institute of Technology, Pasadena CA 91125, USA\\%10
\and
CSIRO Space and Astronomy, PO Box 76, Epping, NSW 1710, Australia\\%11
\and
South-Western Institute For Astronomy Research, Yunnan University, Yunnan 650504, China\\%12
\vs\no\\
   {\small Received 20xx month day; accepted 20xx month day}}
\abstract{As the third paper in the multiple-part series, we report the statistical properties of radio bursts detected from the repeating fast radio burst (FRB) source FRB 20201124A with the Five-hundred-meter Aperture Spherical radio telescope (FAST) during an extremely active episode between the 25th and the 28th of September 2021 (UT). We focus on the polarisation properties of 536 bright bursts with $\mathrm{S/N}>50$. We found that the Faraday rotation measures (RMs) monotonically dropped from $-579 \ {\rm rad \ m^{-2}}$ to $-605 \ {\rm rad \ m^{-2}}$ in the 4-day window. The RM values were compatible with the values ($-300$ to $-900\ {\rm rad \ m^{-2}}$ ) reported 4 month ago \citep{Xuheng2022Nature}. However, the RM evolution rate in the current observation window was at least an order of magnitude smaller than the one ($\sim 500\ {\rm rad \ m^{-2}\, day^{-1}}$) previously reported  during the rapid RM-variation phase, but is still higher than the one ($\le 1\ {\rm rad \ m^{-2} day^{-1}}$ ) during the later RM no-evolution phase. The bursts of FRB~20201124A were highly polarised with the total degree of polarisation (circular plus linear) greater than 90\% for more than 90\% of all bursts. The distribution of linear polarisation position angles (PAs), degree of linear polarisation ($L/I$), and degree of circular polarisation ($V/I$) can be characterised with unimodal distribution functions. During the observation window, the distributions became wider with time, i.e. with larger scatter, but the centroids of the distribution functions remained nearly constant. For individual bursts, significant PA variations (confidence level 5-$\sigma$) were observed in 33\% of all bursts. The polarisation of single pulses seems to follow certain complex trajectories on the Poincar\'e sphere, which may shed light on the radiation mechanism at the source or the plasma properties along the path of FRB propagation.
}
   \authorrunning{Jiang et al.}            %author_head in even pages
   \titlerunning{FAST polarimetry of FRB20201124A}  % title_head in odd pages
   \maketitle
%% The author head (on even pages) and the title head (on odd pages) will be
%% automatically extracted from \author{} and \title{}. Whenever the title is too long,
%% you will be asked to supply a shorter one by inserting either \authorrunning{} or
%% \titlerunning{} before \maketitle. Anyway, you can specify your own heads.
%%
%%
%% Note: In the following text body of your manuscript, please note several differences from
%%       other major journals:
%% (1) \subsection{Please Capitalize the First Letter of Each Notional Word in Subsection Title}
%% (2) Please Capitalize the First Letter of Each Notional Word in all tables' captions

%
%________________________________________________ sections below
%
\section{Introduction}           %% first-level sections will be auto-capitalized
\label{sect:intro}
Radio bursts from a large fraction of fast radio bursts (FRBs) are polarised. The polarisation properties of 9 FRB repeaters have been reported, viz. FRB~20121102A, 20180301A, 20180916B, 20190303A, 20190417A, 20190520B, 20190604A, 20190711A and 20201124A. In general, the radio burst emission is linear polarisation dominated. No significant circular polarisation has been detected except for the FRB~20190520B \citep{2022arXiv220211112A,2022arXiv220308151D} and FRB~20201124A \citep{Xuheng2022Nature,Hilmarsson2021MNRAS,2022MNRAS.512.3400K}. 

FRB~20121102A is the most intensively studied FRB repeater. It was discovered with the Arecibo Telescope \citep{2014ApJ...790..101S}, and later confirmed to be a repeater \citep{2016Natur.531..202S}. Its bursts are $\sim$ 100\% linearly polarised with flat position angle (PA) curves as measured in the C-band \citep{Michilli2018Natur,Gajjar2018ApJ}. However, the degree of linear polarisation becomes significantly smaller in the L-band \citep{Li2021Nature,2022MNRAS.511.6033P,feng2022science}. The Faraday rotation measure (RM) of FRB~20121102A ($\sim 10^5\,\mathrm{rad\,m^{-2}}$) is much larger than other FRBs, which showed a long-term decreasing trend and short-term variations \citep{Michilli2018Natur,Gajjar2018ApJ,Faber2021RNAAS,Hilmarsson2021ApJ}.

FRB~20180301A was discovered by the Parkes ``Murriyang'' radio telescope \citep{2019MNRAS.486.3636P}, and was confirmed to be a repeater with the FAST follow-up observations \citep{Luo2020Natur}. Its bursts were linearly polarised but the degrees of linear polarisation could be significantly less than 100\%, and the PA can be either constant or varying across individual bursts \citep{Luo2020Natur,feng2022science}.

FRB~20180916B was discovered by the Canadian Hydrogen Intensity Mapping Experiment (CHIME) \citep{CHIME2019ApJ}. Its bursts are also $\sim$100\% linearly polarised above 300 MHz with flat PA curves \citep{CHIME2019ApJ,Chawla2020ApJ,Nimmo2021NatAs,2021Natur.596..505P,2021arXiv211102382S}.  The degree of linear polarisation decreases to 30\% -- 70\% at 110 -- 188 MHz \citep{Pleunis2021ApJ}.

FRB~20190520B was discovered with FAST \citep{2021arXiv211007418N}. Most of its detected bursts are linearly polarised with the degree of linear polarisation ranging from 15\% to 80\% \citep{2022arXiv220211112A,2022arXiv220308151D}. A single incidence of FRB~20190520B exhibited $(42\pm 7)\%$ circular polarisation \citep{2022arXiv220211112A}.

FRB~20190711A was discovered by the Australian Square Kilometre Array Pathfinder (ASKAP) with $\sim$100\% linear polarisation and flat PA curves \citep{Day2020MNRAS}. An extremely band-limited repetition detected by Parkes Telescope also showed a high degree of linear polarisation (80\%) and a flat PA \citep{Kumar2021MNRAS}.

Repeating FRBs~20190303A, 20190417A and 20190604A were discovered by CHIME \citep{2020ApJ...891L...6F}. Bursts from FRB~20190303A are 100\% linearly polarised at L-band with flat PA curves, and its RM changed by $\sim 100\,\mathrm{rad\,m^{-2}}$ in 1.5 years \citep{feng2022science}. Bursts from FRB~20190417A are also linearly polarised with flat PA curves, but their degree of linear polarisation ranges from 52\% to 86\% \citep{feng2022science}. A burst from FRB~20190604A exhibited 100\% linear polarisation with a flat PA curve \citep{2020ApJ...891L...6F}.

In this paper, we focus on the polarisation properties of FRB 20201124A. The source was discovered by CHIME. It entered a period of high activity between March and May, 2021 \citep{CHIME2021ATel,Xuheng2022Nature,2022ApJ...927...59L}. During the active episode, the source showed very rich polarisation behaviours unprecedented among other FRB repeaters. \cite{Hilmarsson2021MNRAS} used Effelsberg Telescope to observe FRB~20201124A at 1.36 GHz on April 9, 2021, and obtained flat PA curves but with degrees of circular polarisation up to 20\%. \citet{2022MNRAS.512.3400K} used Parkes Telescope to observe the source and detected a burst with about 47\% of circular polarisation, and a change of about $50^\circ$ in PA was also detected between its two components. The FAST observations detected 1863 bursts from the source \citep{Xuheng2022Nature}. Many interesting features were observed from this large sample of bursts: 1) both PA swings and flat PA profiles are found from the bursts; 2) high degrees of circular polarisation up to 75.1\% have been detected in a fraction of the bursts; and 3) apparent oscillations between linear and circular polarisations are observed in some bursts.

As the third paper in the multiple-part series, this paper reports the polarisation properties of FRB 20201124A observed in its active window between the 25th and the
28th of September 2021 (UT). The current paper is organised as follows. We describes the setups of our observation in Section~\ref{sect:obs}. In Section~\ref{sect:data} the data reduction procedures are explained. The results of all detected bursts are shown and discussed in Section~\ref{sect:results} and conclusions are drawn in Section~\ref{sect:conclusion}. The burst morphology, energetics statistics and timing results for the same set of data are reported in companion papers D. J. Zhou et al. (2022, Paper I), Y. K. Zhang et al. (2022, Paper II), and J. R. Niu et al. (2022, Paper IV), respectively. 

\section{Observation}
\label{sect:obs}
We observed FRB~20201124A with the FAST using the central beam of the 19-beam receiver \citep{8105012,li18IMMag,FASTcommission,FASTperformance}.
%,8073111,8251424,7695900
The source was observed daily between September 25 to 28th, 2021 (UT). The length of each observing session was approximately one hour, and the starting epochs (in MJD) of the four sessions were 59482.942361, 59483.861806, 59484.813194, and 59485.781944, respectively. FAST was pointed to the coordinate provided by the European VLBI Network (EVN), i.e. $\mathrm{RA}=05^\mathrm{h}08^\mathrm{m}03.5077^\mathrm{s}$, $\mathrm{Dec}=+26^\circ 03'38.504''$ \citep{EVN2021ATel14603,2022ApJ...927L...3N}. In the beginning and the end of each observation, 1-minute noise signals were injected for the purpose of polarisation calibration (see Section~\ref{sec:polmty}). We sampled the raw voltage using a \textsc{roach2}-board based digital backend \citep{FASTcommission,FASTperformance}, which recorded the search-mode filterbank data in 8-bit \texttt{PSRFITS} format \citep{PSRCHIVE2004PASA}. We recorded the full coherency matrix, and later converted into the 4-channel Stokes-parameters format. 
For the full bandwidth of 1.0 to 1.5 GHz, our frequency resolution was $4096\, {\rm channels} \times 122\, {\rm kHz}$, and the time resolution was 49.152~$\rm \mu s$.

\section{Data reduction}
\label{sect:data}

\subsection{FRB searching}
We used the software package \texttt{TransientX}\footnote{\url{https://github.com/ypmen/TransientX}} to search off-line for FRB bursts in the data. For FRB~20201124A, the DM and DM fluctuation had been known (${\rm DM}\sim410 \,\mathrm{pc\,cm^{-3}}$ with ${\rm \Delta DM} \sim 10 \,\mathrm{pc\,cm^{-3}}$, \citealt{CHIME2021ATel, Xuheng2022Nature}), so we set the DM search scheme with a uniform grid, of which the DM step is $0.1\,\mathrm{pc\,cm^{-3}}$ and the DM range is $405-420\,\mathrm{pc\,cm^{-3}}$. Burst signals were searched using the matched filters with the equal-SNR-loss grid of time widths ranging from 0.1 to 100
ms \citep{Men2019MNRAS}. 2$\times$20 MHz bandpass edges were removed on both sides to avoid aliasing effect. Following the community convention \citep{2021MNRAS.503.5223Z}, the detection threshold of signal-to-noise ratio (S/N) is chosen to be 7. All the search candidates were later verified by eye to exclude RFI contamination, despite that significant amount of RFIs were automatically mitigated with \texttt{TransientX} using algorithms described in \citet{Men2019MNRAS}.

As discussed in Paper~I, the estimation of dispersion measure (DM) could be affected by the distorted pulse pattern for individual bursts with multi-peak or drifting morphology. Examining DMs of bursts with multiple components showed no systematic variations in the timescale of one hour. Therefore, after the burst detection, we re-perform  dedispersion using the average DM of each observation in Table~1 of Paper~I, i.e. the daily average DMs are $\mathrm{DM_{mean}}=$412.38, 412.25, 412.53, and 411.56 $\mathrm{cm^{-3}\,pc}$, respectively, from September 25 to 28.

Barycentric burst times were converted from topocentric times using \texttt{TEMPO2}\footnote{\url{https://bitbucket.org/psrsoft/tempo2/}} \citep{tempo2_I,tempo2_use,tempo2_software}. Barycentric burst times are in Barycentric Coordinate Time (TCB), and the reference frequency is 1.5 GHz.

In the following polarisation analysis, we raised the threshold of $\mathrm{S/N}$ to 50. There are two major reasons. On one hand, we need to channelise the data to measure the PA rotation as a function of frequency to get RM. The S/N per channel is lower than that of the full band. In this way, RM can be only measured reliably with high S/N data. On the other hand, we need a good S/N in polarisation intensity.  The degree of polarisation could be low (a few percent) for some bursts, which further forces us to raise the S/N threshold of selecting bursts in the polarisation analysis. The threshold of $\rm S/N\ge 50$ cuts down the numbers of bursts to 15, 39, 153, and 329 in the days of MJD 59482 to 59485, respectively. The numbers of bursts included in our analysis are, thus, different from other papers in this series of papers on FAST observations of FRB~20201124A.

\subsection{Polarimetric calibration}
\label{sec:polmty}
The 19-beam receiver of FAST uses orthogonal linear polarisation feeds. At the beginning and the end of each observation, wide-band noise diode signals were injected as the $45^{\rm \circ}$ 100\% linearly polarised calibrator signal. The noise signals were amplitude modulated with a periodic square wave, whose period and duty cycle were 0.201326592 s and 50\%, respectively. We use the software package \texttt{DSPSR}\footnote{\url{http://dspsr.sourceforge.net}} \citep{DSPSR2011PASA} to fold the calibration files, and use the single-axial model of software package \texttt{PSRCHIVE}\footnote{\url{http://psrchive.sourceforge.net}} \citep{PSRCHIVE2004PASA,PSRCHIVE2012AR&T} for polarimetric calibration. In this manuscript, we follow the PSR/IEEE convention \citep{PSRCHIVE2010PASA} for the definition of Stokes parameters, where the total intensity is denoted as $I$, the two linear parameters as $Q,U$,  and the circular component as $V$. Our polarisation presentation on the Poincar\'e sphere is defined by
\begin{eqnarray}
Q&=&P\cos2\chi\cos2\Psi,\\
U&=&P\cos2\chi\sin2\Psi,\\
V&=&P\sin2\chi,
\end{eqnarray}
where $\Psi$ is the linear position angle and $\chi$ is the ellipticity angle. The total polarisation intensity is $P=\sqrt{Q^2+U^2+V^2}$.

Part of the observations on September 27 and 28 (MJD 59484 and 59484) entered the large zenith angle ($>26.4^\circ$), beyond which the effective reflector is not axial symmetric. We have evaluated how such the large zenith angle illumination affects the polarimetry of FAST, the details of which are presented in Appendix~\ref{sec:large_zenith}.

It turns out that the algebraic operations of computing the linear and total polarisation using the Stokes parameters leads to statistical biases. Thus, after the polarisation calibration, we perform the generalised Weisberg correction \citep{Everett2001ApJ} before computing the linear and total polarisation intensities, which are defined as $L=\sqrt{Q^2+U^2}$ and $P=\sqrt{Q^2+U^2+V^2}$. The Weisberg correction modifies the algebraic operation to 
\begin{eqnarray}
L_\mathrm{true}&=&\sqrt{\sum_{i=1,2}S_{i}^2-\varepsilon_L}\,,\\
P_\mathrm{true}&=&\sqrt{\sum_{i=1,2,3}S_{i}^2-\varepsilon_P}\,.
\end{eqnarray}
where $S_1=Q$, $S_2=U$, and $S_3=V$. The subscript is introduced to shorten the notation.
The corrections are
\begin{eqnarray}
\varepsilon_L&=&\frac{\sum_{p(1,2)}S_{p(i)}^2\sigma_{p(j)}^2}{\sum_{i=1,2} S_{i}^2}\,\\
\varepsilon_P&=&\frac{\sum_{p(1,2,3)}(S_{p(i)}^2+S_{p(j)}^2)\sigma_{p(k)}^2}{\sum_{i=1,2,3} S_{i}^2}\,,
\end{eqnarray}
where $\sum_{\rm p(i,j)}$ or $\sum_{\rm p(i,j,k)}$ perform summation over all possible permutation of indices $i,j$ or $i,j,k$.

\subsection{Absolute flux determination}\label{sec:flux_estimation}
We estimate the burst flux from the signal-to-noise ratio (S/N), as no flux calibrator was arranged to save the telescope time. The expected mean flux $S$ is computed from radiometer equation 
\begin{equation}
S_\nu=\frac{T_\mathrm{sys}(\mathrm{S/N})}{G\sqrt{2B\tau}},
\end{equation}
where $T_\mathrm{sys}$ is the effective system temperature, $G$ is the effective gain of telescope, $B$ is the bandwidth used for detection, and $\tau$ is the integration time. The factor 2 comes from combining the two orthogonal polarisations. Here, the system efficiency is already included in the effective gain and system temperature.
The specific fluence ($F_\nu$) was then derived by integrating the specific flux over the pulse width of the FRB burst,
\begin{equation}
    F_\nu=\int_{t_0}^{t_1} S_\nu(\nu, t)\mathrm dt,
\end{equation}
where $t_1-t_0$ is pulse width. The mean fluence ($F$) is the average value of fluence over the pulse signal bandwidth (i.e. $B=\int_{f_1}^{f_2} \mathrm d\nu$, see \citealt{Xuheng2022Nature} for the details of bandwidth definition) that
\begin{equation}
    F=\frac{1}{B}\int_{f_1}^{f_2} F_\nu\mathrm d\nu\,,
\end{equation}
with which the burst energy ($E$) becomes
\begin{equation}
    E=\frac{4\pi D_\mathrm{L}^2}{1+z}FB\,,
\end{equation}
where $D_\mathrm{L}=453.3\,\mathrm{Mpc}$ is the luminosity distance of FRB~20201124A. We used the redshift $z=0.09795$ measured by \citet{Xuheng2022Nature}. 

{In the above estimation for the absolute flux, the \emph{observed bandwidth} of burst signal is defined within the observing frequency window, i.e. 1 to 1.5 GHz. However, the \emph{intrinsic bandwidth} could extend beyond the observing window. Thus, a Gaussian fitting method\citep{Xuheng2022Nature} is used to estimate intrinsic bandwidths, which are shown in Fig.~\ref{fig:properties}, \ref{fig:triplot} and \ref{fig:energy_bandwidth}.}

\section{Results and Discussions}\label{sect:results}

We adopt 9 different parameters to characterise the polarisation properties of the bursts from FRB~20201124A. The parameters are 1) RM, 2) mean position angle $\Psi$, 3) maximum change in the position angle across the pulse profile $\Delta\Psi$, 4) the maximum change in the ellipticity angle across the pulse profile $\Delta\chi$, 5) average degree of linear polarisation $L/I$, 6) average degree of circular polarisation $V/I$, 7) average degree of total polarisation $P/I$, 8) burst energy $E$, and 9) burst bandwidth $B$ using Gaussian fitting. The methods to measure $L/I$,$V/I$, $P/I$, $E$,and $B$ are described in 
Section~\ref{sec:polmty}, \ref{sec:flux_estimation}. %\textcolor{blue}{and \ref{sec:gaussian_bandwidth}}.
The definition and inference of the rest of the parameters are lengthy and technical, which are summarised in Appendix~\ref{sec:faraday_rotation}, \ref{sec:dleta_psi}, \ref{sec:delta_chi}, and \ref{sec:circular_std} for readers' reference.

\begin{figure}[!hbt]
\centering
\includegraphics[width=3.0in]{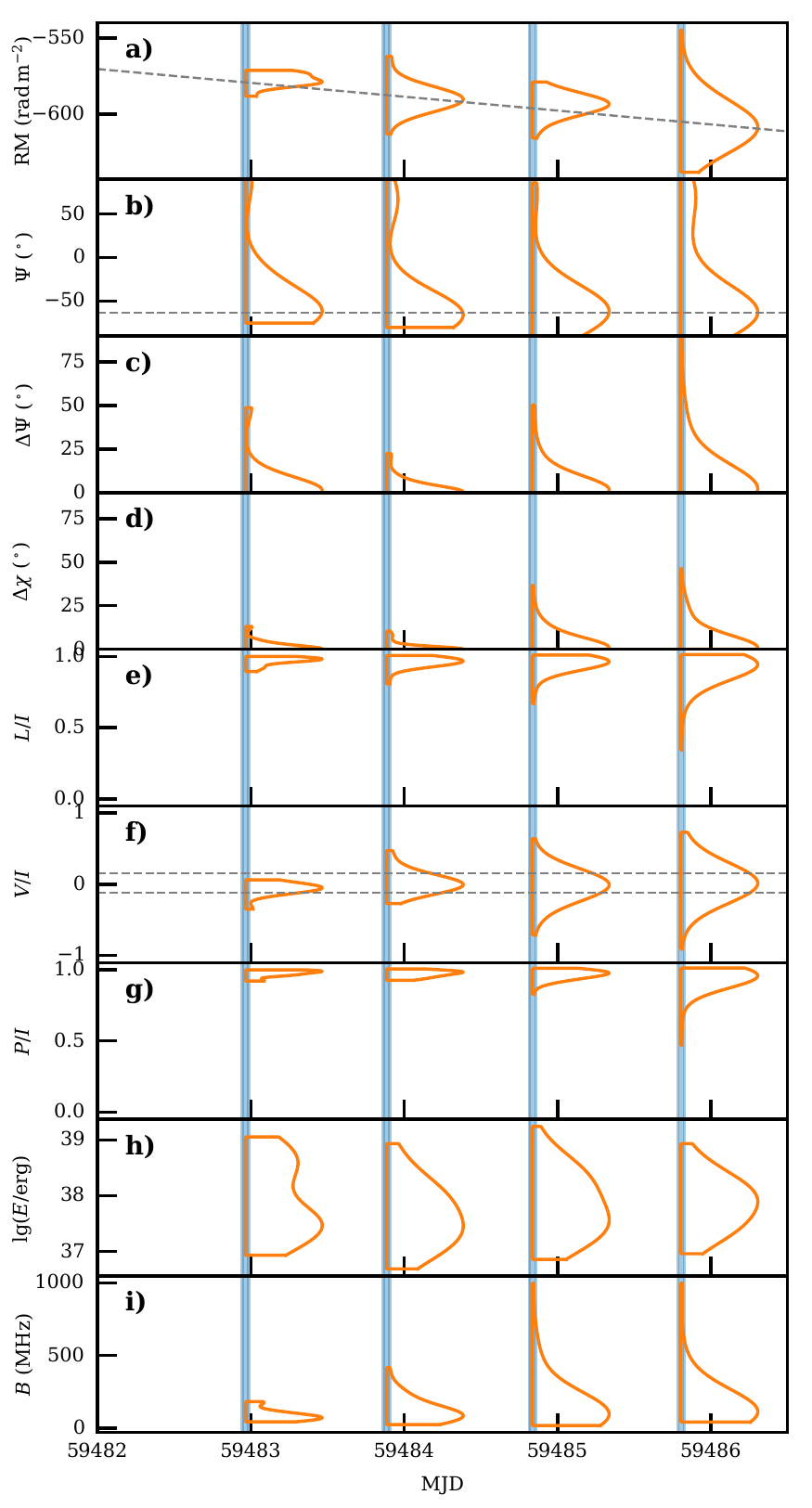}
\caption{Violin plots (orange lines) of the polarisation properties of the bursts. (\textbf{a}) Rotation measure (RM) of interstellar and intergalactic media after subtracting the contribution of the ionosphere of the earth. The dashed grey line is the fitting of a linear model, as a function of time, to all the measure RM values. (\textbf{b}) Mean position angle $\Psi$ of the bursts, corrected to infinite frequency. The dashed grey line shows the averaged value of the position angle  weighted by burst energy. (\textbf{c}) maximum change of position angle $\Psi$ in each burst. (\textbf{d}) maximum change of ellipticity angle $\chi$ in each burst. (\textbf{e}) Degree of linear polarisation $L/I$ after correcting for Faraday rotation. (\textbf{f}) Degree of circular polarisation $V/I$. The two dashed grey lines denote the 68\% interval of the distribution across the four days. (\textbf{g}) Degree of polarisation $P/I$. (\textbf{h}) Burst energy $E$ in the observed bandwidth. (\textbf{i}) Burst bandwidth $B$ in the observed bandwidth. The vertical blue spans denote the observing time ranges.\label{fig:properties}}
\end{figure}

\begin{figure*}[!hbt]
\centering
\includegraphics[width=\linewidth]{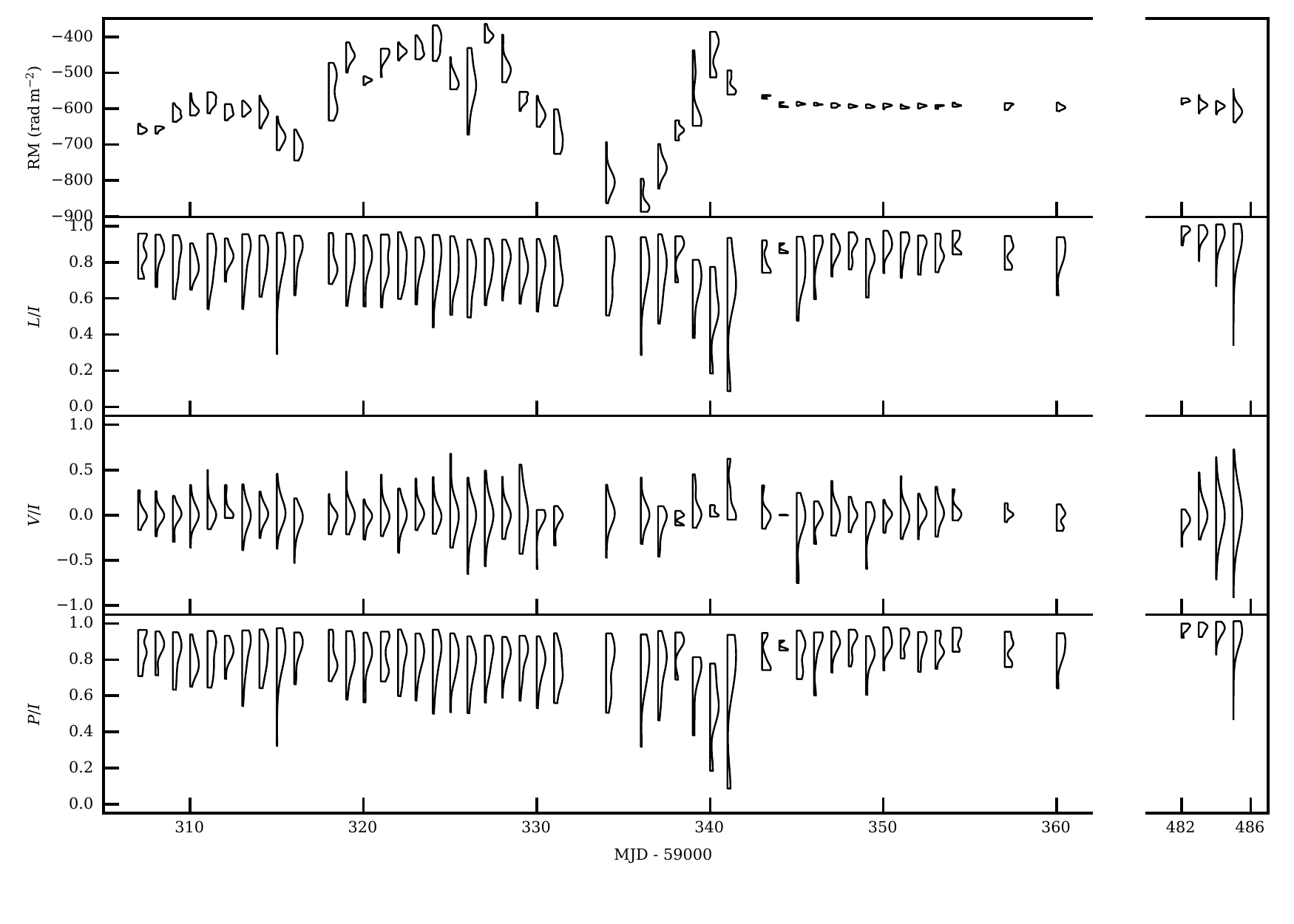}
\caption{Comparison of RM between observations in March to May and in September, 2021. The results between March to May were in \cite{Xuheng2022Nature}.\label{fig:poln_long}}
\end{figure*}

The daily distribution of burst properties are shown in Fig.~\ref{fig:properties} and the related statistics are reported in Table.~\ref{tab:std}. The distributions in Fig.~\ref{fig:properties} are presented using violin plots, where the curve of each epoch shows the distribution function of measured values. The density is estimated using the Gaussian kernel method implemented in \texttt{scikit-learn}\footnote{\url{https://scikit-learn.org/stable/modules/generated/sklearn.neighbors.KernelDensity.html}} \citep{scikit-learn}. The \emph{most probable} values (the values at which the distribution density are maximal) of all parameters show no significant changes except for the RM. The RMs after subtracting the contribution of the earth's ionosphere showed a decreasing rate of $(-9.3\pm 0.5)\,\mathrm{rad\, m^{-2}\,d^{-1}}$ by linearly fitting the value of each burst. After subtracting the linear variation, the Kolmogorov-Smirnov test show that the p-values is smaller than 0.05 to accept the hypothesis that any two data samples are from the same population, except for the first (MJD 59482) and the third day (MJD 59484), where p-value is 0.53. Note that the absolute value of RM, in fact, \emph{increases} due to the fact that $\mathrm{RM}<0$. The evolution of RM over the four-day observations verifies the previous conclusion \cite{Xuheng2022Nature} that there exists a highly variable magnetoionic environment around the FRB source with a spatial scale of $0.5\,{\rm au}\,(v/100\,{\rm km\,s^{-1}}) (\tau/10\,{\rm day})$. The daily RM scatter is increasing, as the distribution becomes wider with time. \citet{Xuheng2022Nature} have shown that the RMs of FRB~20201124A followed the cold plasma Faraday rotation relation that the rotation of polarisation plane was proportional to the square of wavelength; and they have also proven that the daily RM fluctuations were not caused by a profile evolution or apparent RM changes due to intrinsic polarisation structures in pulsed signals. Therefore, we attribute the RM fluctuation to the increase of activities in the FRB local environment, where the RM fluctuation could be caused either by a larger scattering disk or a higher amplitude of the plasma turbulence. We can compare the current observation with the results in May (Fig.~\ref{fig:poln_long}). One can see that the RM values were compatible with what we saw in May. The fluctuation level of daily RM was, however, higher than that in May, but was compatible to the RM fluctuation level seen in March to April \citep{Xuheng2022Nature}. Clearly, certain mechanisms to alter the plasma properties of the FRB environment on the timescale of month are expected.

\begin{table*}[!hbt]
\centering
\caption{Standard deviations of $\Psi$, $\Delta\Psi$, $\Delta\chi$, $L/I$, $V/I$ and $P/I$ of the bursts detected in different days.  Due to the $\pi$ periodicity, $\sigma_\Psi$ is defined with the circular statistics \citep{CircularS} as explained in Appendix~\ref{sec:circular_std}. The rounding of measured values and errors in this table and other parts of the manuscript follow the rounding rules defined at \url{https://pdg.lbl.gov/2021/reviews/rpp2021-rev-rpp-intro.pdf\#page=18}.}
\label{tab:std}
\begin{tabular}{c c c c c}
\hline
MJD & 59482 & 59483 & 59484 & 59485\\
\hline
$\sigma_\Psi$ ($^\circ$) & $18 \pm 6$ & $19.2 \pm 3.5$ & $19 \pm 4$ & $27.1 \pm 1.6$ \\
$\sigma_{\Delta\Psi}$ ($^\circ$) & $12.2 \pm 2.2$ & $5.3 \pm 0.6$ & $9.6 \pm 0.6$ & $13.7 \pm 0.5$ \\
$\sigma_{\Delta\chi}$ ($^\circ$) & $3.4 \pm 0.6$ & $2.54 \pm 0.29$ & $6.2 \pm 0.4$ & $7.72 \pm 0.30$ \\
$\sigma_{(L/I)}$ & $0.029 \pm 0.005$ & $0.038 \pm 0.004$ & $0.0533 \pm 0.0030$ & $0.0841 \pm 0.0033$ \\
$\sigma_{(V/I)}$ & $0.094 \pm 0.017$ & $0.143 \pm 0.016$ & $0.169 \pm 0.010$ & $0.201 \pm 0.008$ \\
$\sigma_{(P/I)}$ & $0.023 \pm 0.004$ & $0.0231 \pm 0.0026$ & $0.0318 \pm 0.0018$ & $0.0625 \pm 0.0024$ \\
\hline
\end{tabular}
\end{table*}

We point out the similarities between the distributions of single burst polarisation properties from FRB~20201124A and of the polarisation properties of pulsar single pulses. As noted by \citet{SCR84}, the pulsar single pulses are highly fluctuating in polarisation properties. Individual pulses can reach a high degree of polarisation, despite that the integrated pulse profiles show a low degree of polarisation. We see similar properties in FRB~20201124A: the individual burst polarisation is also highly fluctuating and the tails in the distributions of $L/I$ and $V/I$ can both extend to extreme values from non-polarised to nearly 100\% fully polarised. However, since we do not find spin-like period here (See Paper IV, Niu et al. 2022), we can not further study the phase dependent polarisation properties at this stage.

The position angle ($\Psi$) tracks down the geometry of the magnetic field lines in the FRB emission region, as the linear polarisation plane is determined by the local magnetic field direction for either the coherent bunched radiation from magnetosphere \citep{Kumar20,Lu20,Wang20,Yang21,Wang22,zhang22} or the synchrotron maser mechanism from magnetic, relativistic shocks far outside the magnetospheres \citep{Metzger19,Beloborodov20,Margalit20}.
We note that the most probable value of the position angles stayed constant, in contrast to what we have seen in RM. In this way, we expect that the geometry of magnetic field lines in the emission region have been maintained with similar configurations in the four-day observation window. This agrees with either picture of the FRB emission region. In the former picture, the geometry is fixed by the strong magnetic fields is the magnetosphere, whereas in the shock wave picture, an upstream ordered magnetic field provides the fixed direction. The overall stable magnetic field structure in emission region is further confirmed by the stability of other polarisation related parameters, including $\Delta \chi$, $\Delta \Psi$, $L/I$,$V/I$, and $P/I$.

On the other hand, the daily standard deviations (i.e. widths of distribution) of $\Psi$, $\Delta \Psi$, $\Delta \chi$, $L/I$, $V/I$, and $P/I$ evolved significantly (more than 5-$\sigma$) in the four-day observations (see the related numbers in Table.~\ref{tab:std}). These effects do not come from the artifacts caused by the increasing number of the detected bursts. In particular, in MJD 59484 and 59485, the numbers of daily detected bursts are more than 150. The statistical fluctuation should introduce errors less than 8\%, which are smaller than the changes in the distribution widths. In the magnetospheric models, the evolution in the distribution width of $\Psi$ indicates that the size of emission region varied such that a variable range of $\Psi$ were covered.

The current observation showed higher degrees of  linear, and total polarisation compared to what we saw in March to May. The average degrees of linear and total polarisation of the current observations were $L/I=(95.8\pm 0.6)\%$, and $P/L=(97.0\pm 0.5)\%$, respectively, while the corresponding values for the the March to May observation were , $(83.4\pm 2.3)\%$, and $(85.4\pm 1.8)\%$, respectively. The average degree of circular polarisation agrees with 0, where $V/I=(-0.5\pm 1.0)\%$ and $(-1.4\pm 1.8)\%$ in the current and the March-to-May observation session, respectively. The distribution of $V/I$,$L/I$ and  $P/I$ also evolved to a more scattered state during the current observation as shown in Fig.~\ref{fig:properties}. Such an evolution was not detected in the previous observation \citep{Xuheng2022Nature} as shown in Fig.~\ref{fig:poln_long}, where the standard deviations of the distribution functions of polarisation degrees are nearly constant, except the possible jitter around 6th May 2021 (MJD 59340).

As the distribution of circular polarisation became more spread out, approximately 0.6\% of bursts in the current observation session had degrees of circular polarisation higher than 70\%. Considering the intrinsic mechanisms, such a high degree of circular polarisation can be produced via the off-beam curvature radiation from bunches \citep{Wang21}, but cannot be achieved for the synchrotron maser mechanism. Propagation effects can also generate circular polarisation through two possible channels: 1) the polarisation dependent radiative transferring \citep{Xuheng2022Nature}, or 2) multi-path propagation \citep{BKN22,feng2022science,Yang2022}. We can exclude neither the intrinsic mechanism nor propagation effects at this stage, because the off-beam curvature radiation can be a natural consequence of an extended radiation region indicated by the scattering of $\Psi$; the propagation mechanisms are supported by the detected polarisation oscillations \citep{Xuheng2022Nature}; and the multi-path propagation may be supported by the increase of the scatter in the RM distribution.

Significant ($>5\sigma$) variations of the position angle across pulse profile ($\Delta\Psi$) were detected in 33\% of the high S/N bursts with $\mathrm{S/N}>50$, while variations of ellipticity angle ($\Delta \chi$) were detected in 28\% of the high S/N bursts. The distribution of $\Delta \chi$ and $\Delta \Psi$ grew wider in the current four-day observation. In the magnetosperic models, variation in distribution width of $\Delta \Psi$ and $\Delta \chi$ can be well understood, if the geometric size of emission region is not stationary as already noticed in the case of pulsars \citep{SCR84}. However, for the shock model, the nonzero values of $\Delta \Psi$ requires a fine-tuned magnetic field configuration to create the swing of polarisation plan during several milliseconds. The nonzero values of $\Delta \chi$ indicate that the degree of circular polarisation changes within the bursts, which could be the consequence when combining nonzero $\Delta \Psi$ and polarisation-dependent radiative transferring\citep{Huang2011MNRAS}.

The distribution functions of energy and signal bandwidth seems to be constant. The 19-beam receiver limits the total bandwidth to 460 MHz (with 20MHz bandedge removed on each side). Such an instrumental limitation introduces artificial sharp cutoffs in the energy distributions in Fig.~\ref{fig:properties}. The widths of either the bandwidth or the energy distribution are wide, that of the burst energies spans for more than two orders of magnitude. Due to the wide widths of the distributions, we cannot conclude if the averages or the widths of bandwidth and energy have changed in our four-day observations.

To understand the physics behind the temporal evolution of parameter distributions, we investigate the correlation between parameter pairs as shown in Fig.~\ref{fig:triplot}.  The zoomed-in version are available in Appendix~\ref{app:zomcor}. For bursts with no obvious (less than 5-$\sigma$) $\Delta \Psi$ or $\Delta \chi$, we plot them with $\Delta \Psi=0$ or $\Delta \chi=0$, which forms line-like artificial structures in the plots. We see clear correlations induced by the parameter dependence, i.e. $L/I$-$V/I$, $L/I$-$P/I$, and energy-bandwidth. Here, the $L/I$-$V/I$ and $L/I$-$P/I$ correlations come from the definition of the total polarisation ($P^2\equiv L^2+V^2$), and the fact that total polarisation is less than the intensity ($P/L\le1$). 

No obvious correlation is found between the RM and the position angle $\Psi$. In this way, we conclude that 1) RM change was not caused by the change in $\Psi$; and 2) the $\Psi$ distribution was intrinsic and not induced by fluctuations in RM. We can attribute the RM change to the FRB environment and $\Psi$ distribution to the FRB intrinsic properties. A self-consistent picture would be that 1) the magnetoionic environment of the FRB is highly variable to produce RM variations, while 2) the geometric center of the emission region is  kept constant to produce a stable average value of $\Psi$. In the same time, 3) the spatial size of emission region is growing larger to produce the increasing fluctuations in the $\Psi$ distribution.

There may be marginal correlations between parameter pairs of $\Delta \Psi$-$L/I$, $\Delta \Psi$-$P/I$, $\Delta \chi$-$L/I$, and $\Delta \chi$-$P/I$. Those correlations are expected, since the degree of linear polarisation would decrease, if the variation of the position angle increases, i.e. degree of polarisation would decrease if one adds up radiation with different position angles. 

There is a weak $\Delta \Psi$--$\Delta \chi$ correlation (see Fig.~\ref{fig:deltapa_deltachi}), where the correlation coefficient is $\rho=0.45$ after excluding points with $\Delta \Psi=0$ and $\Delta \chi=0$. This indicates polarisation dependent radiative transfer, in which Stokes parameters can convert from one to another\footnote{For interested readers, the polarisation dependent radiative transferring in the context of FRB is addressed in \citet{Xuheng2022Nature}, which includes the Faraday conversion discussed later in \citet{KS22}}. We investigate the phenomenon by checking the phase-resolved relation between the two parameters as shown in Fig.~\ref{fig:poincare_example} (more examples are given in Appendix~\ref{sec:poincare}). We can see the the correlation between $\Psi$ and $\chi$ along the burst time and clear patterns in the trajectories of phase-resolved polarisation profile on the surface of Poincar\'e sphere (panel (c) of Fig.~\ref{fig:poincare_example}). The polarisation trajectories are of a great diversity. They can be a straight line or a single curve (Burst 738), or a curve with branches (Burst 237), or a more complex shape (Burst 437). If such trajectories were generated at the pulse emission stage, a complex geometry for the magnetospheric magnetic fields is required. On the other hand, if it is it due to polarisation radiative transfer, we expect that the magnetic field in the plasma medium is time dependent and also follows a simple geometric configuration. 

Certain aspects of the polarisation behaviours can be understood within the scenario of bunched coherent curvature radiation, which allows for a diversity in polarisation. The charged bunches can produce considerably circular polarisation, when the line of sight sweeps across the rim of the radiation beam (the angular size $\sim 1/\gamma$); while emissions are strongly linearly polarised, when the opening angle of the bunch is much larger than $1/\gamma$ \citep{Wang22}. Furthermore, different kinds of evolutionary trajectories can be reproduced on the Poincar{\'e} sphere \citep{Wang21}. Our results suggest that most bunches have large opening angles to produce high degree of linear polarisation, while a small fraction of the bunches have smaller opening angles and display an off-axial generation of radio waves.

\begin{figure*}[!hbt]
\centering
\includegraphics[width=\linewidth]{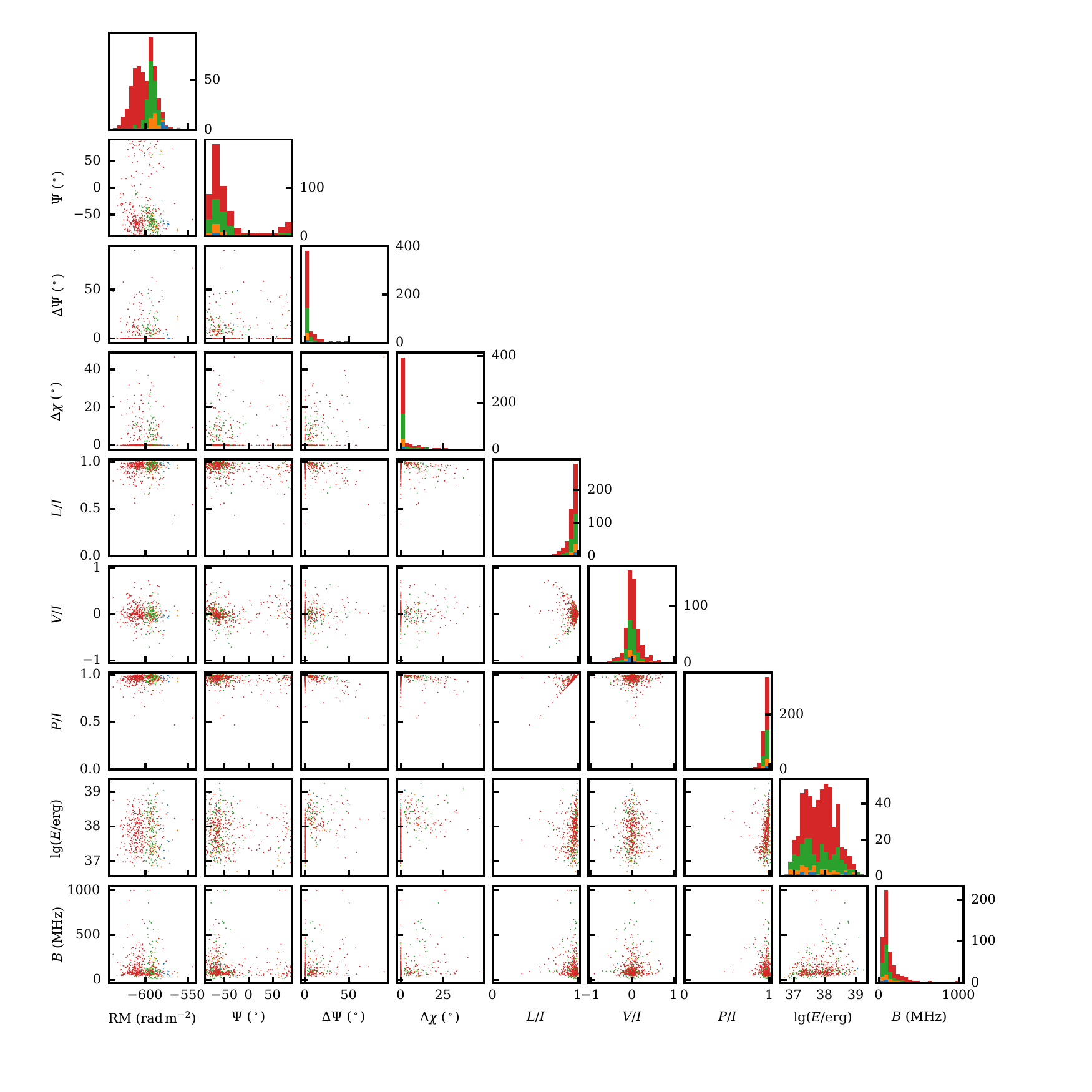}
\caption{The scatterplot matrix of DM, RM, position angle $\Psi$, maximum change of position angle $\Delta\Psi$, degree of linear polarisation $L/I$, degree of circular polarisation $V/I$, burst energy $E$ and burst bandwidth $B$. Bursts on MJD 59482, 59483, 59484 and 59485 are marked in blue, orange, green and red respectively.}
\label{fig:triplot}
\end{figure*}

\begin{figure}[!hbt]
\centering
\includegraphics[width=0.24\linewidth]{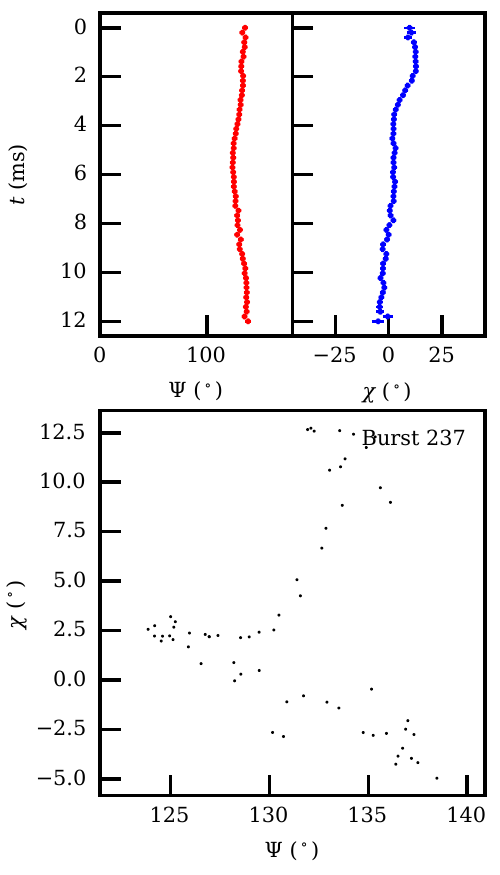}
\includegraphics[width=0.24\linewidth]{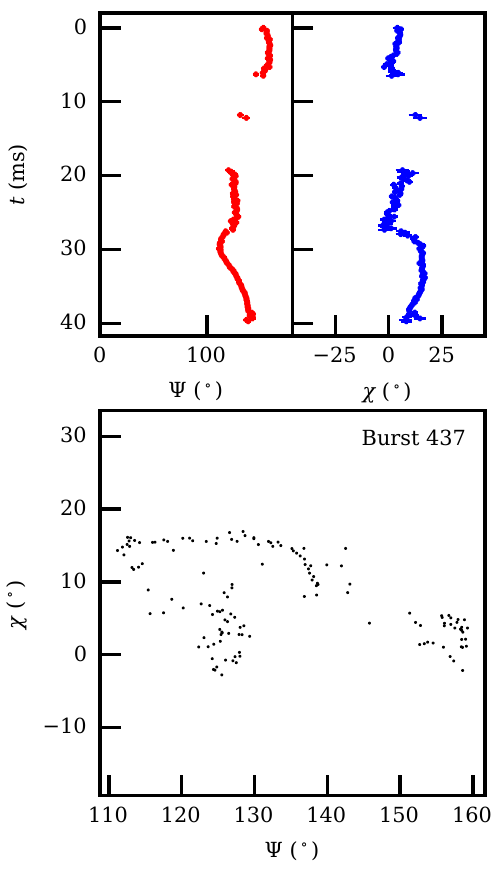}
\includegraphics[width=0.24\linewidth]{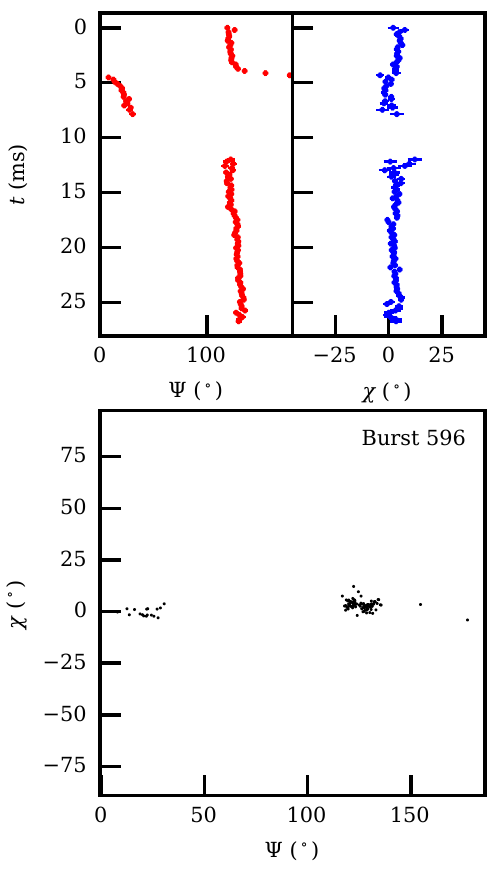}
\includegraphics[width=0.24\linewidth]{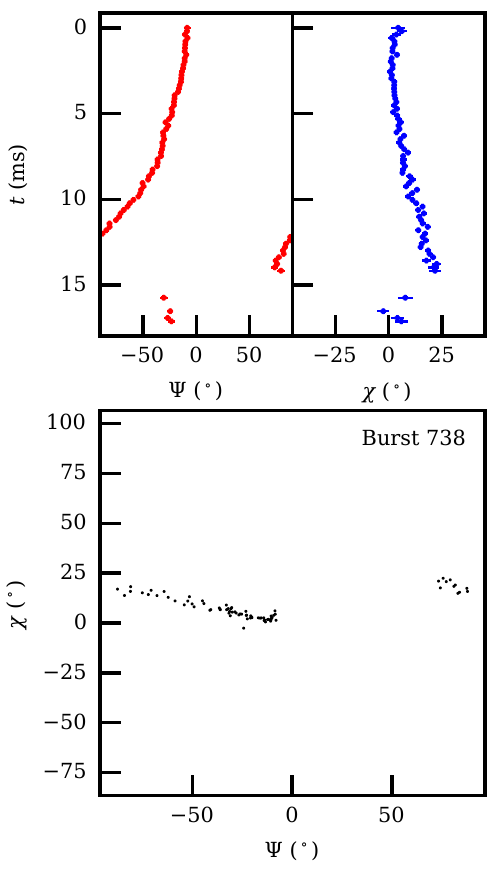}
\caption{$\Psi$ and $\chi$ curves of 4 bursts. (\textbf{a}) The curve of position angle $\Psi$. (\textbf{b}) The curve of ellipticity angle $\chi$. (\textbf{c}) Correlation between $\Psi$ and $\chi$ curves. Error bars in (\textbf{a}) and (\textbf{b}) denote 68\% confidence level.\label{fig:poincare_example}}
\end{figure}

\section{Conclusions}
\label{sect:conclusion}
In this manuscript, we presented the statistical results of FAST polarimetric observations of FRB~20201124A from 25th to 28th of September 2021. We also compared the current results with the previous observations carried out half a year ago. We detected RM evolution and pulse-to-pulse RM scatter with properties similar to previous results. 

We note that the diversity in the burst polarisation properties is similar to the case of radio pulsars, which disfavours the relativistic shock-wave models for FRBs and provide further evidence for the magnetospheric origin of FRBs. 
The most-probable values of the polarisation degrees, i.e. $V/I$, $L/I$, and $P/I$, are stable throughout the observation period, but the values are slightly higher than seen a couple of months ago. We note that the distribution of the polarisation degrees is more scattered than that of the March to May session. 

We have excluded that the RM variations cause the evolution of the polarisation properties, and we are able to isolate the environment contribution from the FRB intrinsic evolution. The distributions of the position angle and polarisation degrees evolved towards a more scattered state within the four-day window. We argue that the phenomena are caused by the expansion of the size of the radiation region in the magnetosphere. 

We found a weak correlation between the position angle variation and the ellipticity angle variation. Such a correlation, together with the complex trajectories of the polarisation profiles in the Poincar\'e sphere representation, suggests that polarisation dependent radiative transfer plays a significant role in shaping the polarisation properties of FRB. 

We note that the ellipticity angle variation correlated with the burst energy. So far there is no theoretical prediction or explanation to this correlation. We expect that a future collection of a larger sample in FRB~20201124A may help to identify or falsify such a correlation. Investigations of the circular polarisation properties of some bursts may help to understand the radiation mechanism of FRBs and the immediate environment of the FRB source through propagation effects.

\begin{acknowledgements}
We thank the anonymous referee for providing valuable suggestions. 
This work made use of the data from FAST (Five-hundred-meter Aperture Spherical radio Telescope). FAST is a Chinese national mega-science facility, operated by National Astronomical Observatories, Chinese Academy of Sciences. 
This is work is supported by the National SKA Program of
China (2020SKA0120100, 2020SKA0120200), the National Key R\&D Program of China (2017YFA0402602), the 
National Nature Science Foundation grant No. 12041303, the
CAS-MPG LEGACY project, and funding from the Max-Planck Partner Group.
J.~L. Han is supported by the National Natural Science Foundation of
China (NSFC, Nos. 11988101 and 11833009) and the Key Research
Program of the Chinese Academy of Sciences (Grant No. QYZDJ-SSW-SLH021);
D.~J. Zhou is supported by the Cultivation Project for the FAST scientific Payoff and Research Achievement of CAMS-CAS.
Y. Feng is supported by the Key Research Project of Zhejiang Lab no. 2021PE0AC0.
Y.P.Y. is supported by National Natural Science Foundation of China grant No. 12003028 and the China Manned Spaced Project (CMS-CSST-2021-B11).

\end{acknowledgements}
\section*{Authors contributions}  
J.~C. Jiang, K.~J. Lee, and B.~Zhang led the charge of writing this paper and performed the polarimetry and data analysis in this paper.
B. Zhang initially proposed the FRB key science project.
B. Zhang, K.~J. Lee, J.~L. Han, W. ~W. Zhu and D. Li coordinated the teamwork, the observational campaign, co-supervised data analyses and interpretations.
W.~Y. Wang, K.~J. Lee, D.~Z. Li, R.~X.Xu, and Y.P.Yang provided theoretical inputs.
D.~J. Zhou studied burts morphology and classification with the details presented in paper I of this series.
Y.~K. Zhang analyzed the energy distribution with the details presented in paper II of this series.
J.~R. Niu performed periodicity search with the results presented in paper IV.
H. Xu, J.~W.~Xu, C.~F. Zhang, B.~J. Wang, D.~J. Zhou, Y. Feng, W.~C.Jing, R. Luo, C.~C Miao, C.~H. Niu, P. Wang, Z.~L. Yang, M. Yuan did the data analysis.
Z.~G. Dai, J.~M. Yao, C.~W. Tsai and F.~Y Wang participated in the joint discussion.

\bibliographystyle{raa}
\bibliography{reference}

\begin{thebibliography}{67}
\providecommand\natexlab[1]{#1}
\providecommand\JournalTitle[1]{#1}

\bibitem[{Anna-Thomas} {et~al.}(2022)]{2022arXiv220211112A}
{Anna-Thomas}, R., {Connor}, L., {Burke-Spolaor}, S., {et~al.} 2022, arXiv
  e-prints, arXiv:2202.11112

\bibitem[{Beloborodov}(2020)]{Beloborodov20}
{Beloborodov}, A.~M. 2020, \apj, 896, 142

\bibitem[{Beniamini} {et~al.}(2022)]{BKN22}
{Beniamini}, P., {Kumar}, P., \& {Narayan}, R. 2022, \mnras, 510, 4654

\bibitem[{Buchner} {et~al.}(2014)]{2014A&A...564A.125B}
{Buchner}, J., {Georgakakis}, A., {Nandra}, K., {et~al.} 2014, \aap, 564, A125

\bibitem[{Chawla} {et~al.}(2020)]{Chawla2020ApJ}
{Chawla}, P., {Andersen}, B.~C., {Bhardwaj}, M., {et~al.} 2020, \apjl, 896, L41

\bibitem[{CHIME/FRB Collaboration} {et~al.}(2019)]{CHIME2019ApJ}
{CHIME/FRB Collaboration}, {Andersen}, B.~C., {Bandura}, K., {et~al.} 2019,
  \apjl, 885, L24

\bibitem[{Chime/Frb Collabortion}(2021)]{CHIME2021ATel}
{Chime/Frb Collabortion}. 2021, The Astronomer's Telegram, 14497, 1

\bibitem[{Dai} {et~al.}(2022)]{2022arXiv220308151D}
{Dai}, S., {Feng}, Y., {Yang}, Y.~P., {et~al.} 2022, arXiv e-prints,
  arXiv:2203.08151

\bibitem[{Day} {et~al.}(2020)]{Day2020MNRAS}
{Day}, C.~K., {Deller}, A.~T., {Shannon}, R.~M., {et~al.} 2020, \mnras, 497,
  3335

\bibitem[{Desvignes} {et~al.}(2019)]{Desvignes2019Sci}
{Desvignes}, G., {Kramer}, M., {Lee}, K., {et~al.} 2019, Science, 365, 1013

\bibitem[Dunning {et~al.}(2017)]{8105012}
Dunning, A., Bowen, M., Castillo, S., {et~al.} 2017, in 2017 XXXIInd General
  Assembly and Scientific Symposium of the International Union of Radio Science
  (URSI GASS), 1

\bibitem[{Efron}(1982)]{Efron82}
{Efron}, B. 1982, {The Jackknife, the Bootstrap and other resampling plans}
  ({CBMS-NSF Regional Conference Series in Applied Mathematics, Society for
  Industrial and Applied Mathematics (SIAM), Philadelphia US})

\bibitem[{Everett} \& {Weisberg}(2001)]{Everett2001ApJ}
{Everett}, J.~E., \& {Weisberg}, J.~M. 2001, \apj, 553, 341

\bibitem[{Faber} {et~al.}(2021)]{Faber2021RNAAS}
{Faber}, J.~T., {Gajjar}, V., {Siemion}, A. P.~V., {et~al.} 2021, Research
  Notes of the American Astronomical Society, 5, 17

\bibitem[{Feng} {et~al.}(2022)]{feng2022science}
{Feng}, Y., {Li}, D., {Yang}, Y.-P., {et~al.} 2022, Science, 375, 1266

\bibitem[{Fisher}(1996)]{CircularS}
{Fisher}, N.~I. 1996, {Statistical Analysis of Circular Data} ({Cambridge
  University Press, Cambridge UK})

\bibitem[{Fonseca} {et~al.}(2020)]{2020ApJ...891L...6F}
{Fonseca}, E., {Andersen}, B.~C., {Bhardwaj}, M., {et~al.} 2020, \apjl, 891, L6

\bibitem[{Gajjar} {et~al.}(2018)]{Gajjar2018ApJ}
{Gajjar}, V., {Siemion}, A.~P.~V., {Price}, D.~C., {et~al.} 2018, \apj, 863, 2

\bibitem[{Hilmarsson} {et~al.}(2021{\natexlab{a}})]{Hilmarsson2021MNRAS}
{Hilmarsson}, G.~H., {Spitler}, L.~G., {Main}, R.~A., \& {Li}, D.~Z.
  2021{\natexlab{a}}, \mnras, 508, 5354

\bibitem[{Hilmarsson} {et~al.}(2021{\natexlab{b}})]{Hilmarsson2021ApJ}
{Hilmarsson}, G.~H., {Michilli}, D., {Spitler}, L.~G., {et~al.}
  2021{\natexlab{b}}, \apjl, 908, L10

\bibitem[{Hobbs}(2012)]{tempo2_use}
{Hobbs}, G. 2012, arXiv e-prints, arXiv:1205.6273

\bibitem[{Hobbs} {et~al.}(2006)]{tempo2_I}
{Hobbs}, G.~B., {Edwards}, R.~T., \& {Manchester}, R.~N. 2006, \mnras, 369, 655

\bibitem[{Hobbs} \& {Edwards}(2012)]{tempo2_software}
{Hobbs}, G., \& {Edwards}, R. 2012, {Tempo2: Pulsar Timing Package}

\bibitem[{Hotan} {et~al.}(2004)]{PSRCHIVE2004PASA}
{Hotan}, A.~W., {van Straten}, W., \& {Manchester}, R.~N. 2004, \pasa, 21, 302

\bibitem[{Huang} \& {Shcherbakov}(2011)]{Huang2011MNRAS}
{Huang}, L., \& {Shcherbakov}, R.~V. 2011, \mnras, 416, 2574

\bibitem[{Jiang} {et~al.}(2019)]{FASTcommission}
{Jiang}, P., {Yue}, Y., {Gan}, H., {et~al.} 2019, Science China Physics,
  Mechanics, and Astronomy, 62, 959502

\bibitem[{Jiang} {et~al.}(2020)]{FASTperformance}
{Jiang}, P., {Tang}, N.-Y., {Hou}, L.-G., {et~al.} 2020, Research in Astronomy
  and Astrophysics, 20, 064

\bibitem[{Kumar} \& {Bo{\v{s}}njak}(2020)]{Kumar20}
{Kumar}, P., \& {Bo{\v{s}}njak}, {\v{Z}}. 2020, \mnras, 494, 2385

\bibitem[{Kumar} {et~al.}(2022{\natexlab{a}})]{2022MNRAS.512.3400K}
{Kumar}, P., {Shannon}, R.~M., {Lower}, M.~E., {et~al.} 2022{\natexlab{a}},
  \mnras, 512, 3400

\bibitem[{Kumar} {et~al.}(2022{\natexlab{b}})]{KS22}
{Kumar}, P., {Shannon}, R.~M., {Lower}, M.~E., {Deller}, A.~T., \& {Prochaska},
  J.~X. 2022{\natexlab{b}}, arXiv e-prints, arXiv:2204.10816

\bibitem[{Kumar} {et~al.}(2021)]{Kumar2021MNRAS}
{Kumar}, P., {Shannon}, R.~M., {Flynn}, C., {et~al.} 2021, \mnras, 500, 2525

\bibitem[{Lanman} {et~al.}(2022)]{2022ApJ...927...59L}
{Lanman}, A.~E., {Andersen}, B.~C., {Chawla}, P., {et~al.} 2022, \apj, 927, 59

\bibitem[{Lee}(2016)]{CPTA2016}
{Lee}, K.~J. 2016, in Astronomical Society of the Pacific Conference Series,
  Vol. 502, Frontiers in Radio Astronomy and FAST Early Sciences Symposium
  2015, ed. L.~{Qain} \& D.~{Li}, 19

\bibitem[{Li} {et~al.}(2018)]{li18IMMag}
{Li}, D., {Wang}, P., {Qian}, L., {et~al.} 2018, IEEE Microwave Magazine, 19,
  112

\bibitem[{Li} {et~al.}(2021)]{Li2021Nature}
{Li}, D., {Wang}, P., {Zhu}, W.~W., {et~al.} 2021, \nat, 598, 267

\bibitem[{Lu} {et~al.}(2020)]{Lu20}
{Lu}, W., {Kumar}, P., \& {Zhang}, B. 2020, \mnras, 498, 1397

\bibitem[{Luo} {et~al.}(2020)]{Luo2020Natur}
{Luo}, R., {Wang}, B.~J., {Men}, Y.~P., {et~al.} 2020, \nat, 586, 693

\bibitem[{Marcote} {et~al.}(2021)]{EVN2021ATel14603}
{Marcote}, B., {Kirsten}, F., {Hessels}, J.~W.~T., {et~al.} 2021, The
  Astronomer's Telegram, 14603, 1

\bibitem[{Margalit} {et~al.}(2020)]{Margalit20}
{Margalit}, B., {Beniamini}, P., {Sridhar}, N., \& {Metzger}, B.~D. 2020,
  \apjl, 899, L27

\bibitem[{Men} {et~al.}(2019)]{Men2019MNRAS}
{Men}, Y.~P., {Luo}, R., {Chen}, M.~Z., {et~al.} 2019, \mnras, 488, 3957

\bibitem[{Metzger} {et~al.}(2019)]{Metzger19}
{Metzger}, B.~D., {Margalit}, B., \& {Sironi}, L. 2019, \mnras, 485, 4091

\bibitem[{Michilli} {et~al.}(2018)]{Michilli2018Natur}
{Michilli}, D., {Seymour}, A., {Hessels}, J.~W.~T., {et~al.} 2018, \nat, 553,
  182

\bibitem[{Nimmo} {et~al.}(2021)]{Nimmo2021NatAs}
{Nimmo}, K., {Hessels}, J.~W.~T., {Keimpema}, A., {et~al.} 2021, Nature
  Astronomy, 5, 594

\bibitem[{Nimmo} {et~al.}(2022)]{2022ApJ...927L...3N}
{Nimmo}, K., {Hewitt}, D.~M., {Hessels}, J.~W.~T., {et~al.} 2022, \apjl, 927,
  L3

\bibitem[{Niu} {et~al.}(2021)]{2021arXiv211007418N}
{Niu}, C.~H., {Aggarwal}, K., {Li}, D., {et~al.} 2021, arXiv e-prints,
  arXiv:2110.07418

\bibitem[{Pastor-Marazuela} {et~al.}(2021)]{2021Natur.596..505P}
{Pastor-Marazuela}, I., {Connor}, L., {van Leeuwen}, J., {et~al.} 2021, \nat,
  596, 505

\bibitem[Pedregosa {et~al.}(2011)]{scikit-learn}
Pedregosa, F., Varoquaux, G., Gramfort, A., {et~al.} 2011, Journal of Machine
  Learning Research, 12, 2825

\bibitem[{Plavin} {et~al.}(2022)]{2022MNRAS.511.6033P}
{Plavin}, A., {Paragi}, Z., {Marcote}, B., {et~al.} 2022, \mnras, 511, 6033

\bibitem[{Pleunis} {et~al.}(2021)]{Pleunis2021ApJ}
{Pleunis}, Z., {Michilli}, D., {Bassa}, C.~G., {et~al.} 2021, \apjl, 911, L3

\bibitem[{Price} {et~al.}(2019)]{2019MNRAS.486.3636P}
{Price}, D.~C., {Foster}, G., {Geyer}, M., {et~al.} 2019, \mnras, 486, 3636

\bibitem[{Robishaw} \& {Heiles}(2021)]{2021hai1.book..127R}
{Robishaw}, T., \& {Heiles}, C. 2021, in The WSPC Handbook of Astronomical
  Instrumentation, ed. A.~{Wolszczan}, 127

\bibitem[{Sand} {et~al.}(2021)]{2021arXiv211102382S}
{Sand}, K.~R., {Faber}, J., {Gajjar}, V., {et~al.} 2021, arXiv e-prints,
  arXiv:2111.02382

\bibitem[{Sotomayor-Beltran} {et~al.}(2013)]{ionfr}
{Sotomayor-Beltran}, C., {Sobey}, C., {Hessels}, J.~W.~T., {et~al.} 2013, \aap,
  552, A58

\bibitem[{Spitler} {et~al.}(2014)]{2014ApJ...790..101S}
{Spitler}, L.~G., {Cordes}, J.~M., {Hessels}, J.~W.~T., {et~al.} 2014, \apj,
  790, 101

\bibitem[{Spitler} {et~al.}(2016)]{2016Natur.531..202S}
{Spitler}, L.~G., {Scholz}, P., {Hessels}, J.~W.~T., {et~al.} 2016, \nat, 531,
  202

\bibitem[{Stinebring} {et~al.}(1984)]{SCR84}
{Stinebring}, D.~R., {Cordes}, J.~M., {Rankin}, J.~M., {Weisberg}, J.~M., \&
  {Boriakoff}, V. 1984, \apjs, 55, 247

\bibitem[{van Straten} \& {Bailes}(2011)]{DSPSR2011PASA}
{van Straten}, W., \& {Bailes}, M. 2011, \pasa, 28, 1

\bibitem[{van Straten} {et~al.}(2012)]{PSRCHIVE2012AR&T}
{van Straten}, W., {Demorest}, P., \& {Oslowski}, S. 2012, Astronomical
  Research and Technology, 9, 237

\bibitem[{van Straten} {et~al.}(2010)]{PSRCHIVE2010PASA}
{van Straten}, W., {Manchester}, R.~N., {Johnston}, S., \& {Reynolds}, J.~E.
  2010, \pasa, 27, 104

\bibitem[{Wang} {et~al.}(2022{\natexlab{a}})]{Wang21}
{Wang}, W.-Y., {Jiang}, J.-C., {Lu}, J., {et~al.} 2022{\natexlab{a}}, Science
  China Physics, Mechanics, and Astronomy, 65, 289511

\bibitem[{Wang} {et~al.}(2020)]{Wang20}
{Wang}, W.-Y., {Xu}, R., \& {Chen}, X. 2020, \apj, 899, 109

\bibitem[{Wang} {et~al.}(2022{\natexlab{b}})]{Wang22}
{Wang}, W.-Y., {Yang}, Y.-P., {Niu}, C.-H., {Xu}, R., \& {Zhang}, B.
  2022{\natexlab{b}}, \apj, 927, 105

\bibitem[{Xu} {et~al.}(2022)]{Xuheng2022Nature}
{Xu}, H., {Niu}, J.~R., {Chen}, P., {et~al.} 2022, \nat, 609, 685

\bibitem[{Yang} {et~al.}(2022)]{Yang2022}
{Yang}, Y.-P., {Lu}, W., {Feng}, Y., {Zhang}, B., \& {Li}, D. 2022, \apjl, 928,
  L16

\bibitem[{Yang} \& {Zhang}(2021)]{Yang21}
{Yang}, Y.-P., \& {Zhang}, B. 2021, \apj, 919, 89

\bibitem[{Zhang}(2022)]{zhang22}
{Zhang}, B. 2022, \apj, 925, 53

\bibitem[{Zhang} {et~al.}(2021)]{2021MNRAS.503.5223Z}
{Zhang}, C.~F., {Xu}, J.~W., {Men}, Y.~P., {et~al.} 2021, \mnras, 503, 5223

\end{thebibliography}

\appendix
\section{FAST polarimetric test at large zenith angles}\label{sec:large_zenith}

The active reflector of FAST consists of 4450 movable reflective panels, which are controlled by actuators to form the instantaneous 300-metre-aperture parabolic surface from a 500-meter aperture spherical surface \citep{FASTcommission}. When the zenith angle reaches $26.4^\circ$, the edge of paraboloid reaches the edge of the spherical reflector, and the paraboloid become asymmetric. Such a large-angle illumination generates systematic error in polarimetry. In addition, the receiver is tilted back towards the reflector surface to avoid stray illuminating coming out of the reflector. The asymmetric reflector and tilted feed illumination generate the beam squint effect \citep{2021hai1.book..127R}, therefore we test polarimetric properties for a large zenith angle observation.

\begin{figure}
    \centering
    \includegraphics[width=65mm]{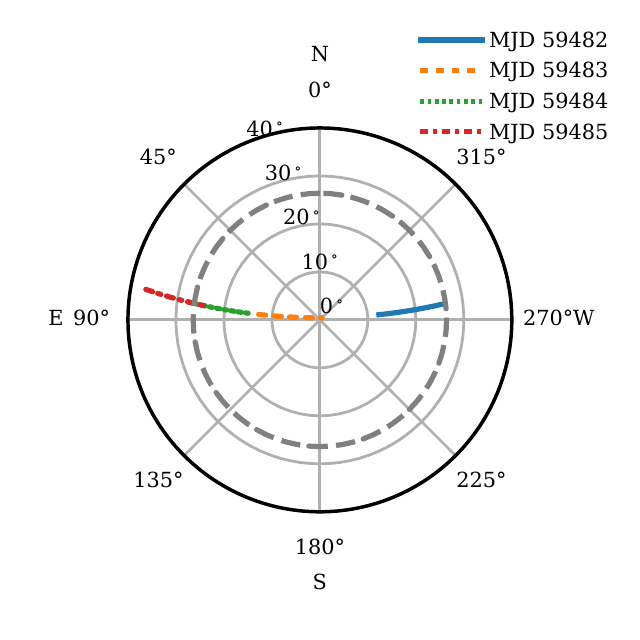}
    \caption{Horizontal coordinates of FRB~20201124A in FAST observations between September 26 (MJD 59482) and 29 (MJD 59485). In the polar plot, the polar angle is the azimuth angle, and the distance is the zenith angle of the source. The dashed grey circle denotes the $26.4^\circ$ limit between small and large zenith angles of FAST. At the end of the observation on September 28 (MJD 59484) and most part of the observation on September 29 (MJD 59485), the zenith angles of the source exceeded the limit.}
    \label{fig:zenith}
\end{figure}

Trajectory of FRB~20201124A in FAST sky coverage is shown in Fig.~\ref{fig:zenith}. At the start of the observation on September 27 (MJD 59484) and most part of the observation on September 28 (MJD 59485), the zenith angles of the source exceeded the $26.4^\circ$ limit of full reflector illumination \footnote{\url{https://fastwww.china-vo.org/cms/article/24/}}. 

\begin{figure}
\centering
\includegraphics[width=65mm]{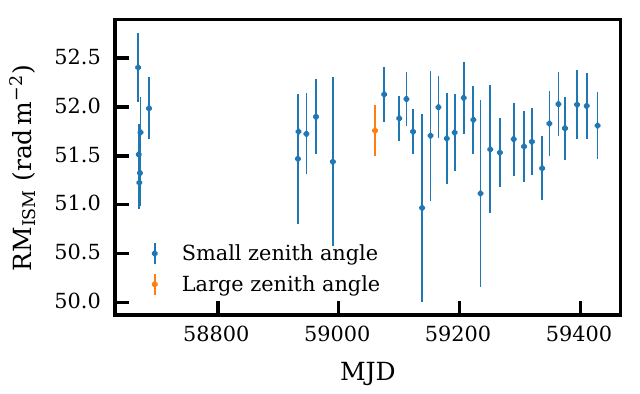}
\caption{Rotation measures of PSR J0621+1002 in CPTA data. The contribution of the ionosphere of the earth has been subtracted. The blue markers denote observations at small zenith angles ($<26.4^\circ$), while the orange marker denote the observation at large zenith angle. Error bars denote 68\% confidence level.}
\label{fig:J0621+1002_rm}
\end{figure}

In order to test FAST polarimetry at large zenith angles ($>26.4^\circ$), we utilised archival data of Chinese Pulsar Timing Array (CPTA) \citep{CPTA2016} and compared the polarimetric results of PSR J0621+1002 at zenith angles smaller and larger than $26.4^\circ$. The test data were obtained between July, 2019 and August, 2021. 35 observations of PSR J0621+1002 were conducted at small zenith angles ($<26.4^\circ$), but the zenith angle of one observation on July 30, 2020 (MJD 59060) was between $29.5^\circ$ and $35.7^\circ$. We used a phase-resolved Bayesian method to fit $\mathrm{RM_{obs}}$ of all observations, and subtracted RM of the ionosphere of the earth $\mathrm{RM_{ion}}$ using \texttt{ionFR}\footnote{\url{https://sourceforge.net/projects/ionfarrot/}} \citep{ionfr} to get the RM of interstellar medium $\mathrm{RM_{ISM}}$. The results in Fig.~\ref{fig:J0621+1002_rm} show that RM is not affected at the level of $0.3\, {\rm rad\,m^{-2}}$ in the large-angle observation.

\begin{figure}
    \centering
    \includegraphics[width=65mm]{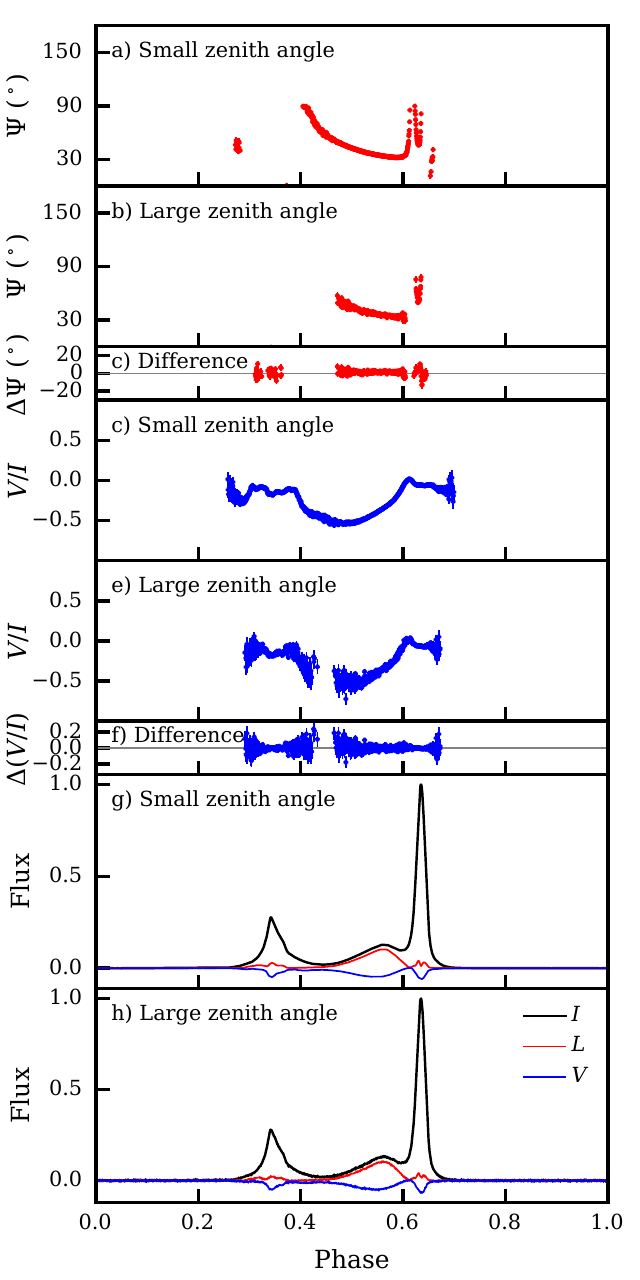}
    \caption{Polarimetric results of PSR J0621+1002. (\textbf{a}) Position angle $\Psi$ at small zenith angle. (\textbf{b}) Position angle $\Psi$ at large zenith angle. (\textbf{c}) Differences in position angle $\Psi$ between large and small zenith angle. (\textbf{d}) Degree of circular polarisation $V/I$ at small zenith angle. (\textbf{e}) Degree of circular polarisation $V/I$ at small zenith angle. (\textbf{f}) Differences in Degree of circular polarisation $V/I$ between large and small zenith angle. (\textbf{g}) polarisation profiles at small zenith angle. (\textbf{h}) polarisation profiles at large zenith angle. Profiles in Panel (\textbf{g}) and (\textbf{h}) are normalised by the peak of total intensity $I$. The black curve is total intensity $I$, the red curve is linearly polarised intensity $L$, and the blue curve is circularly polarised intensity $V$. Error bars in Panel (\textbf{a)} to (\textbf{f}) denote 68\% confidence level.}
    \label{fig:J0621+1002}
\end{figure}

We also combined the polarisation profiles of small zenith angle observations after correcting Faraday rotation using $\mathrm{RM_{obs}}$ of each observation, and compared the combined profile with the polarisation profile at large zenith angle. The results shown in Fig.~\ref{fig:J0621+1002} show no significant difference in polarisation profiles except signal-to-noise ratio because the total observation time at small zenith angle was much longer than the observation at large zenith angle. The differences in position angle between the observations of large and small zenith angles was $1.4^\circ\pm 0.8^\circ$. For the degree of circular polarisation $|V|/I$, the result of combined small zenith angle observations is $(16.282\pm0.010)\%$, and the result of the large zenith angle observation is $(16.92\pm0.07)\%$, therefore the systematic error of $|V|/I$ is less than 1\%. We conclude that at the 1\% level, the polarimetry is not affected by the large-angle illumination.

\section{Measure RM}\label{sec:faraday_rotation}
\begin{figure}
    \centering
    \includegraphics[width=0.7\linewidth]{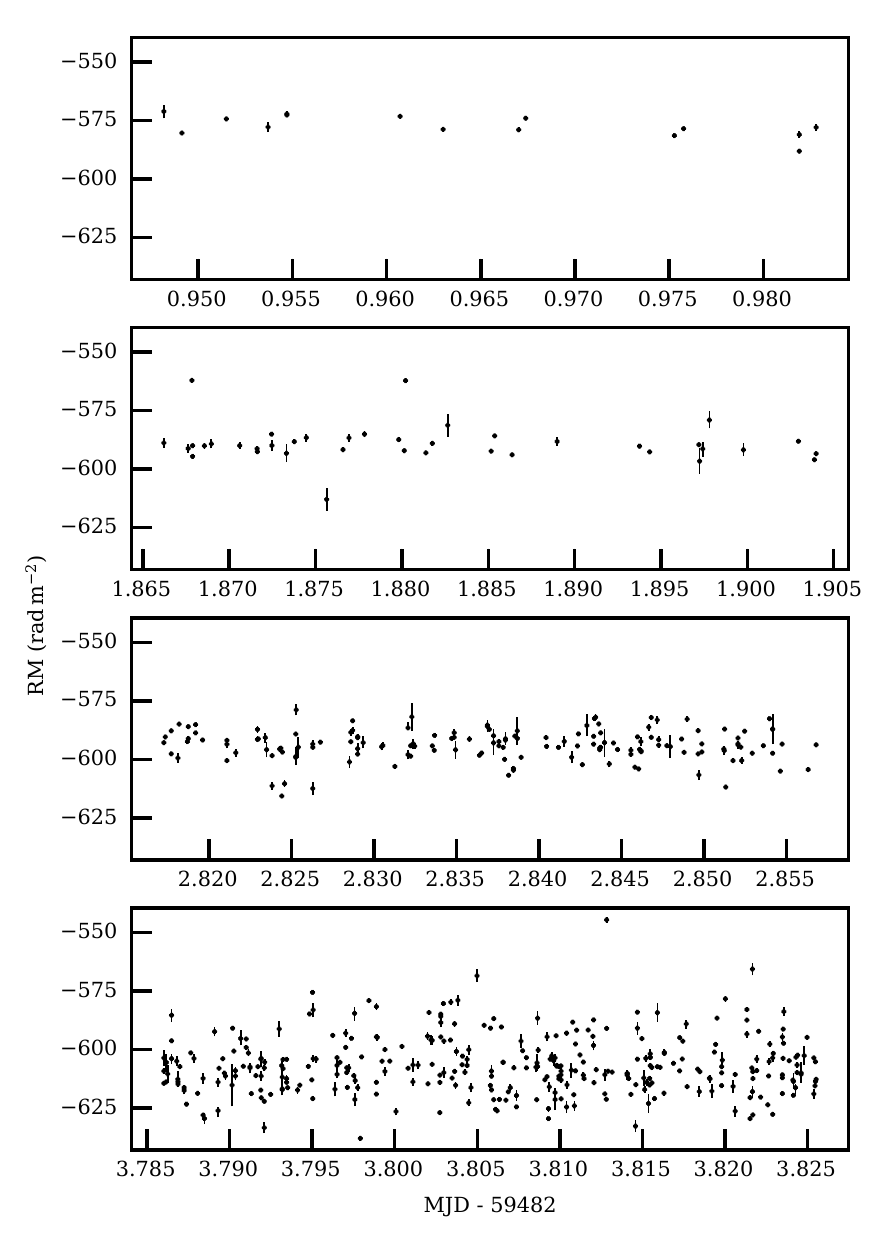}
    \caption{Example of fitted RMs for bursts in September, 2021. The errorbars denote 68\% confidence intervals. Four panels are used to show observing sessions.}
    \label{fig:rm_errorbar}
\end{figure}

When linearly polarised radio wave propagates in the cold magnetized plasma, the plane of polarisation rotate. Such rotation is called Faraday rotation. The angle of polarisation plane, i.e. the position angle $\Psi$, is wavelength ($\lambda$) dependent that 
\begin{equation}
    \Delta\Psi=\mathrm{RM}\,\lambda^2\,,
\end{equation}
where the coefficient $\rm RM$ is the Faraday rotation measure. We used a Bayesian method \citep{Desvignes2019Sci,Luo2020Natur} to fit the rotation in term of Stokes $Q$ and $U$ of the burst to obtain the RM. The best fit value and uncertainty interval for the RM of each burst were derived from the posterior inference using \texttt{PyMultiNest}\footnote{\url{https://johannesbuchner.github.io/PyMultiNest/}}  \citep{2014A&A...564A.125B}. More details of the fitting method can be found in the Supplementary Meterials of \citet{Desvignes2019Sci}. The ionosphere of the earth also contributes to the observed RMs. Software packagee \texttt{ionFR}\footnote{\url{https://sourceforge.net/projects/ionfarrot/}} \citep{ionfr} was used to correct the RM contribution due to the ionosphere and geomagnetic field. The results are shown in Fig.~\ref{fig:rm_errorbar}.

\section{Maximal $\Delta \Psi$}\label{sec:dleta_psi}
The difference of position angle $\Delta\Psi$ can be calculated using Stokes $Q$ and $U$.
\begin{eqnarray}
\Delta\Psi_{ij}&=&\frac{1}{2}\arccos D_{\Psi,ij}\,,\\
D_{\Psi, ij}&=&\frac{Q_iQ_j+U_iU_j}{\sqrt{(Q_i^2+U_i^2)(Q_j^2+U_j^2)}}\,,
\end{eqnarray}
where subscript $i$ and $j$ denote that the $\Delta \Psi_{ij}$ is the difference of position angle between the $i$-th and the $j$-th sample of a given pulse profile.
The uncertainty in $\Delta\Psi$ is computed with error propagation formula that
\begin{equation}
\sigma_{\Delta\Psi,ij}=\frac{1}{2}\sqrt{\frac{\sigma_{D_{\Psi,ij}}^2}{1-D_{\Psi, ij}^2}}\,,
\end{equation}
with 
\begin{equation}
    \begin{aligned}
\sigma_{D_{\Psi,ij}}^2=&\left[\left(\frac{\partial D_\Psi}{\partial Q_i}\right)^2+\left(\frac{\partial D_\Psi}{\partial Q_j}\right)^2\right]\sigma_Q^2\\
&+\left[\left(\frac{\partial D_\Psi}{\partial U_i}\right)^2+\left(\frac{\partial D_\Psi}{\partial U_j}\right)^2\right]\sigma_U^2
\end{aligned}
\end{equation}
and 
\begin{eqnarray}
\frac{\partial D_\Psi}{\partial Q_i}&=&\frac{U_i(Q_jU_i-Q_iU_j)}{L_i^3L_j},\\
\frac{\partial D_\Psi}{\partial Q_j}&=&\frac{U_j(Q_iU_j-Q_jU_i)}{L_j^3L_i},\\
\frac{\partial D_\Psi}{\partial U_i}&=&\frac{Q_i(Q_iU_j-Q_jU_i)}{L_i^3L_j},\\
\frac{\partial D_\Psi}{\partial U_j}&=&\frac{Q_j(Q_jU_i-Q_iU_j)}{L_j^3L_i}.
\end{eqnarray}
Here the error for $Q$ and $U$, i.e. $\sigma_Q$ and $\sigma_U$, are estimated using the standard deviation of $Q$ and $U$ in the off-burst baselines.
The maximal position angle variation can thus be defined as
\begin{equation}
\Delta \Psi=\max_{\forall i,j\,{\rm s.t.}\, \frac{\Delta\Psi_{ij}}{\sigma_{\Delta\Psi,ij}}\ge5 }  \Delta \Psi_{ij}\,.
\end{equation}
When condition $\Delta\Psi/\sigma_{\Delta\Psi}>5$ could not be met ($\sim66$\% of bursts with $\mathrm{S/N}>50$), we denoted $\Delta\Psi$ as 0.

\section{Maximal $\Delta \chi$}
\label{sec:delta_chi}
Similar to the case of position angle $\Delta\Psi$, the difference of ellipticity angle $\Delta\chi$ between the $i$-th and $j$-th data points is.
\begin{eqnarray}
&&\begin{aligned}
D_{\chi,ij}&=\frac{L_iL_j+V_iV_j}{P_i P_j}\\
&=\frac{\sqrt{(Q_i^2+U_i^2)(Q_j^2+U_j^2)}+V_iV_j}{\sqrt{(Q_i^2+U_i^2+V_i^2)(Q_j^2+U_j^2+V_j^2)}},\\
\end{aligned}\\
&&\Delta\Psi_{ij}=\frac{1}{2}\arccos D_{\chi,ij}.
\end{eqnarray}
%where $L_i=\sqrt{Q_i^2+U_i^2}$ and $P_i=\sqrt{Q_i^2+U_i^2+V_i^2}$. 
The uncertainty in $\Delta\chi$ is given by
\begin{eqnarray}
&&\frac{\partial D_\chi}{\partial Q_i}=\frac{Q_iV_i(L_jV_i-L_iV_j)}{L_iP_i^3P_j},\\
&&\frac{\partial D_\chi}{\partial Q_j}=\frac{Q_jV_j(L_iV_j-L_jV_i)}{L_jP_j^3P_i},\\
&&\frac{\partial D_\chi}{\partial U_i}=\frac{U_iV_i(L_jV_i-L_iV_j)}{L_iP_i^3P_j},\\
&&\frac{\partial D_\chi}{\partial U_j}=\frac{U_jV_j(L_iV_j-L_jV_i)}{L_jP_j^3P_i},\\
&&\frac{\partial D_\chi}{\partial V_i}=\frac{L_i(L_iV_j-L_jV_i)}{P_i^3P_j},\\
&&\frac{\partial D_\chi}{\partial V_j}=\frac{L_j(L_jV_i-L_iV_j)}{P_j^3P_i},\\
&&\begin{aligned}
\sigma_{D_\chi}^2=&\left[\left(\frac{\partial D_\chi}{\partial Q_i}\right)^2+\left(\frac{\partial D_\chi}{\partial Q_j}\right)^2\right]\sigma_Q^2\\
&+\left[\left(\frac{\partial D_\chi}{\partial U_i}\right)^2+\left(\frac{\partial D_\chi}{\partial U_j}\right)^2\right]\sigma_U^2\\
&+\left[\left(\frac{\partial D_\chi}{\partial V_i}\right)^2+\left(\frac{\partial D_\chi}{\partial V_j}\right)^2\right]\sigma_V^2
\end{aligned}\\
&&\sigma_{\Delta\chi}=\frac{1}{2}\sqrt{\frac{\sigma_{D_\chi}^2}{1-D_\chi^2}}.
\end{eqnarray}
The maximal value of $\Delta \chi$ is
\begin{equation}
\Delta \chi=\max_{\forall i,j\,{\rm s.t.}\, \frac{\Delta\chi_{ij}}{\sigma_{\Delta\chi,ij}}\ge5 }  \Delta \chi_{ij}\,.
\end{equation}
In our observation, 72\% bursts with $\mathrm{S/N}>50$ can not meet the condition $\Delta\chi/\sigma_{\Delta\chi}>5$, they are marked as $\Delta\chi=0$.

\section{Circular standard deviation of position angle}\label{sec:circular_std}

Due to a periodic boundary of $180^{\circ}$, the circular standard deviation of position angle $\sigma_\Psi$ is defined
\begin{eqnarray}
\bar{C}&=&\frac{1}{N}\sum_{i=0}^{N-1}\cos 2\Psi_i\,,\\
\bar{S}&=&\frac{1}{N}\sum_{i=0}^{N-1}\sin 2\Psi_i\,,\\
\bar{R}&=&\sqrt{\bar{C}^2+\bar{S}^2}\,,\\
\sigma_\Psi&=&\frac{1}{2}\sqrt{-2\ln\bar{R}}\,,
\end{eqnarray}
where $\Psi_i$ is the position angle of each burst, and $N$ is the number of bursts.

The error of $\sigma_\Psi$ is inferred using the standard re-sampling technique \citep{Efron82}, where the values of each day is randomly re-sampled without putting back. The circular standard deviation is estimated by re-scaling the post-sampling ensemble standard deviation with the ensemble size factor.

\section{Zoomed-in correlation relation}
\label{app:zomcor}
\begin{figure}[!hbt]
\centering
\includegraphics[width=0.5\linewidth]{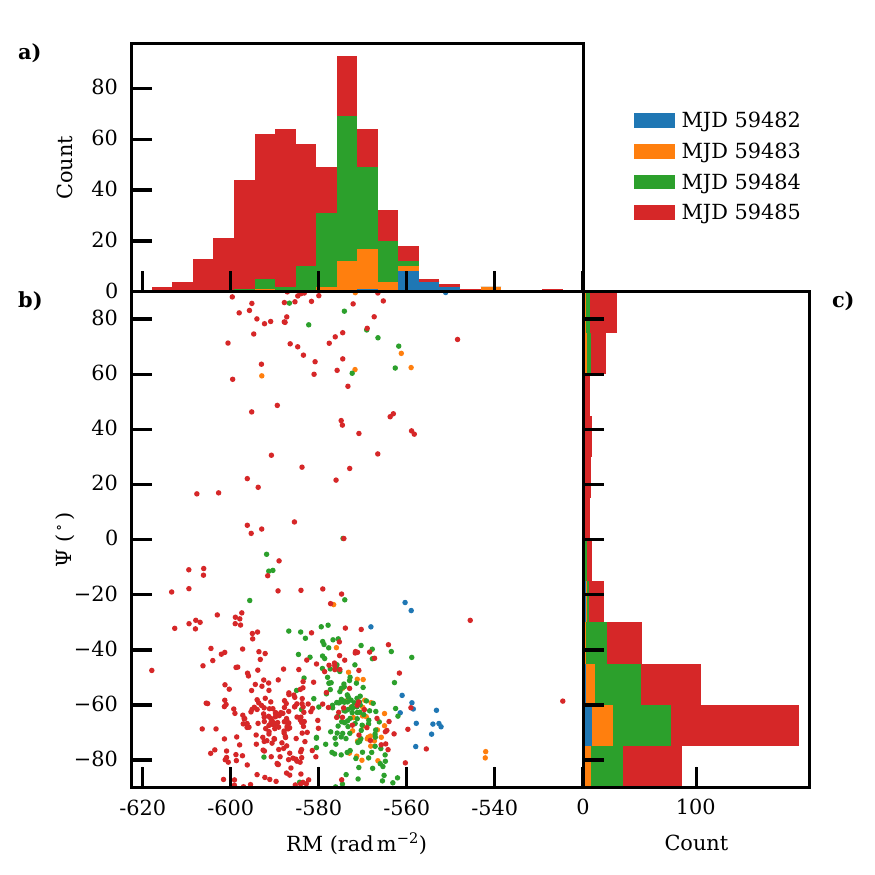}
\caption{(\textbf{a}) Histogram of RM. (\textbf{b}) Correlation between RM and PA. (\textbf{c}) Histogram of position angle $\Psi$.}
\label{fig:rm_pa}
\end{figure}

\begin{figure}[!hbt]
\centering
\includegraphics[width=0.5\linewidth]{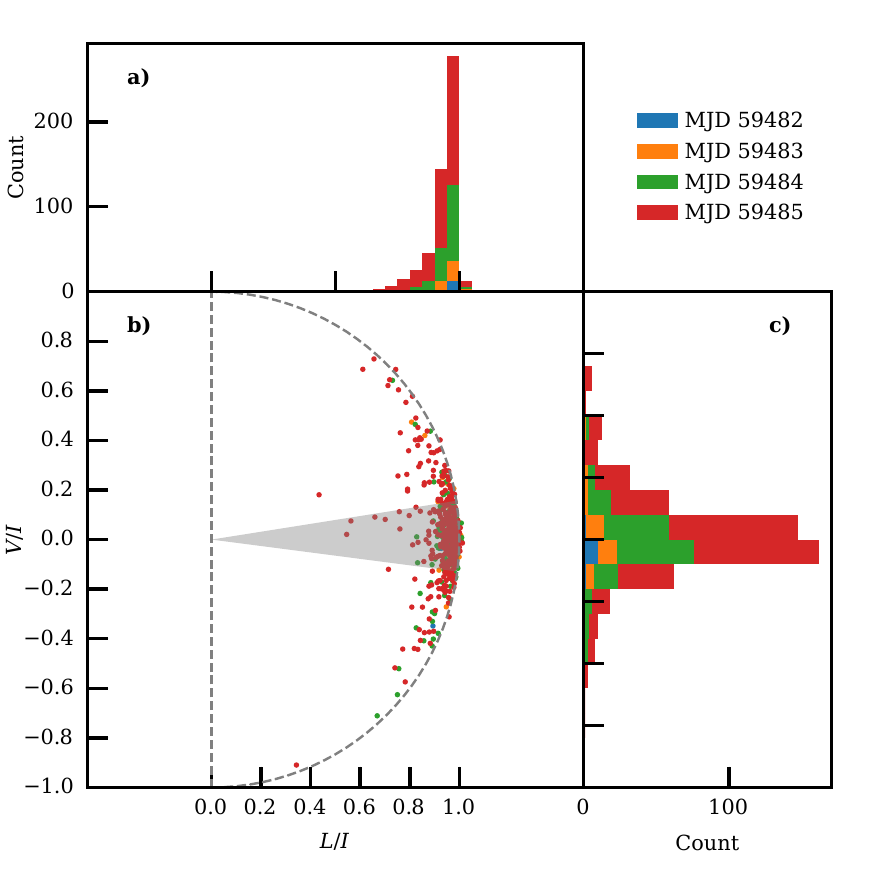}
\caption{(\textbf{a}) Histogram of degree of linear polarisation $L/I$. (\textbf{b}) Correlation between degree of linear polarisation $L/I$ and degree of circular polarisation $V/I$. The grey dashed semicircle denotes the area in which the degree of polarisation $P/I=\sqrt{L^2+V^2}/I\leq1$. The grey sector covers 68\% of bursts. (\textbf{c}) Histogram of degree of circular polarisation $V/I$.}
\label{fig:L_V}
\end{figure}

\begin{figure}[!hbt]
\centering
\includegraphics[width=0.5\linewidth]{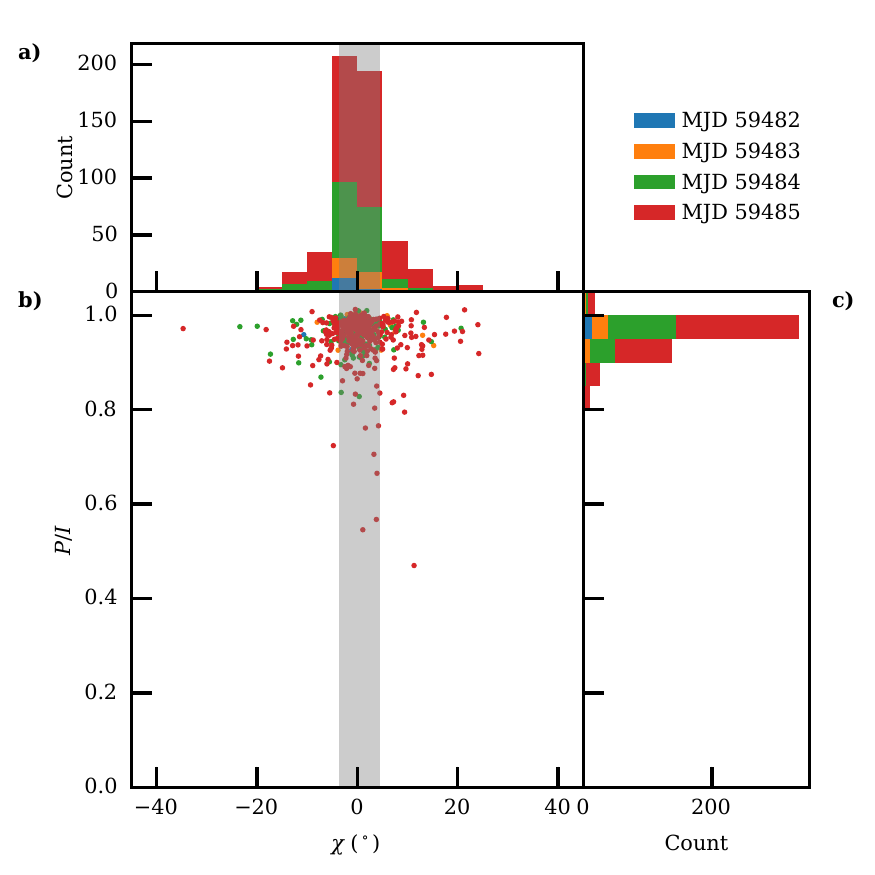}
\caption{(\textbf{a}) Histogram of ellipticity angle $\chi$. (\textbf{b}) Correlation between ellipticity angle $\chi$ and degree of polarisation $P/I$. (\textbf{c}) Histogram of degree of polarisation $P/I$. The grey shade covers 68\% of bursts.}
\label{fig:chi_dop}
\end{figure}

\begin{figure}[!hbt]
\centering
\includegraphics[width=0.5\linewidth]{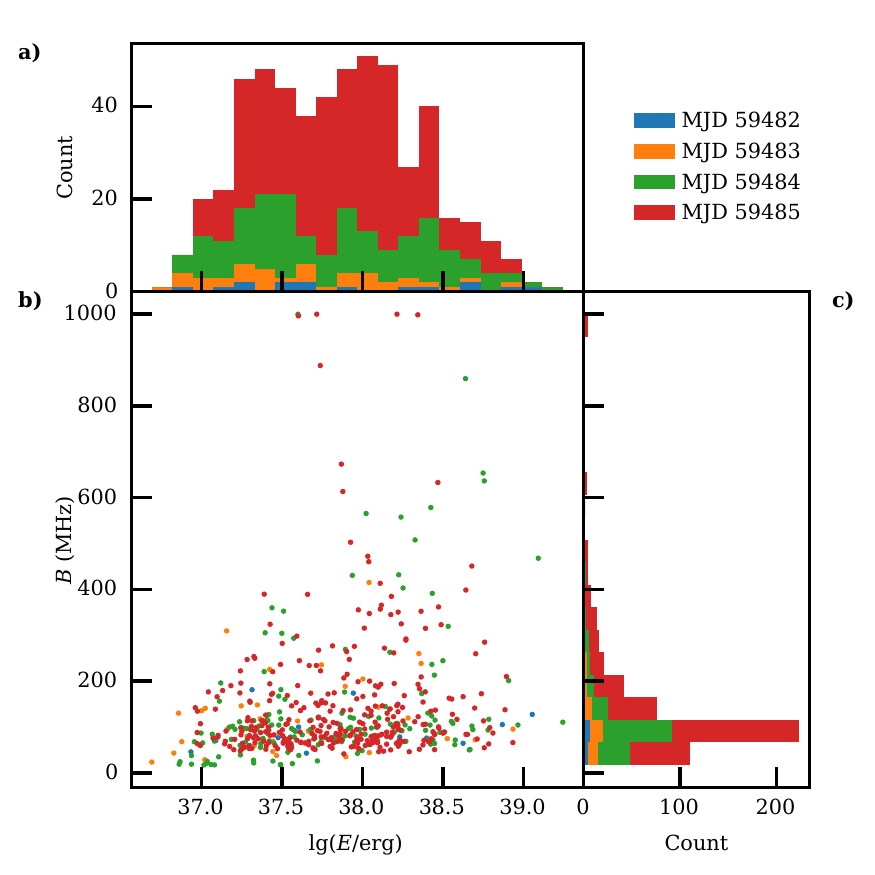}
\caption{(\textbf{a}) Histogram of burst energy $E$. (\textbf{b}) Correlation between burst energy $E$ and burst bandwidth $B$. (\textbf{c}) Histogram of burst bandwidth $B$.}
\label{fig:energy_bandwidth}
\end{figure}

In this section, we collect the zoomed-in version of Fig.~\ref{fig:triplot}, where the RM-$\Psi$, $L$-$V$, $\chi$-$P/I$, $L/I$-$V/I$, and $E$-$B$ correlations are in Fig.~\ref{fig:rm_pa} to \ref{fig:energy_bandwidth}.

\begin{figure}[!hbt]
\centering
\includegraphics[width=0.5\linewidth]{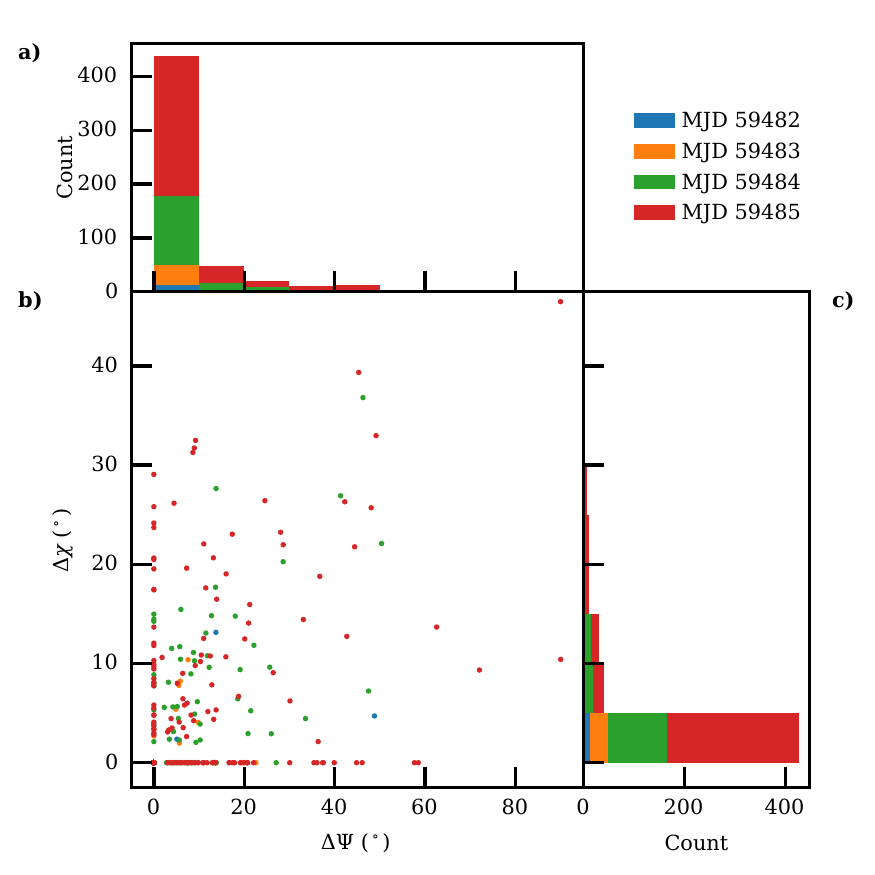}
\caption{(\textbf{a}) Histogram of the maximum change in the position angle across the pulse profile $\Delta\Psi$. (\textbf{b}) Correlation between $\Delta\Psi$ and $\Delta\chi$. (\textbf{c}) Histogram of the maximum change in the ellipticity angle across the pulse profile $\Delta\chi$.}
\label{fig:deltapa_deltachi}
\end{figure}

\section{$\Psi$ and $\chi$ curves}\label{sec:poincare}
Fig.~\ref{fig:poincare_all} collect the phase resolved $\Psi$-$\chi$ curves of all bursts with $\Delta\Psi >20^\circ$, $\Delta\chi >10^\circ$ and $\mathrm{S/N}>200$.
\begin{figure*}
\centering
\includegraphics[width=0.19\linewidth]{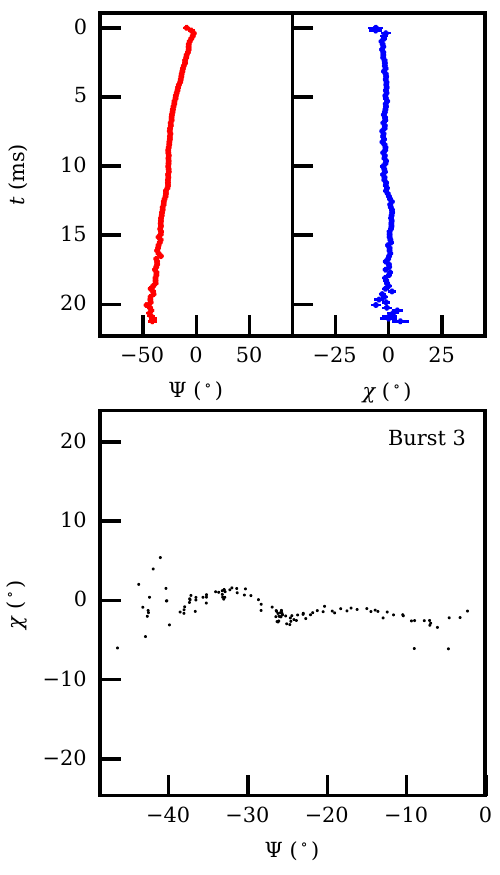}
\includegraphics[width=0.19\linewidth]{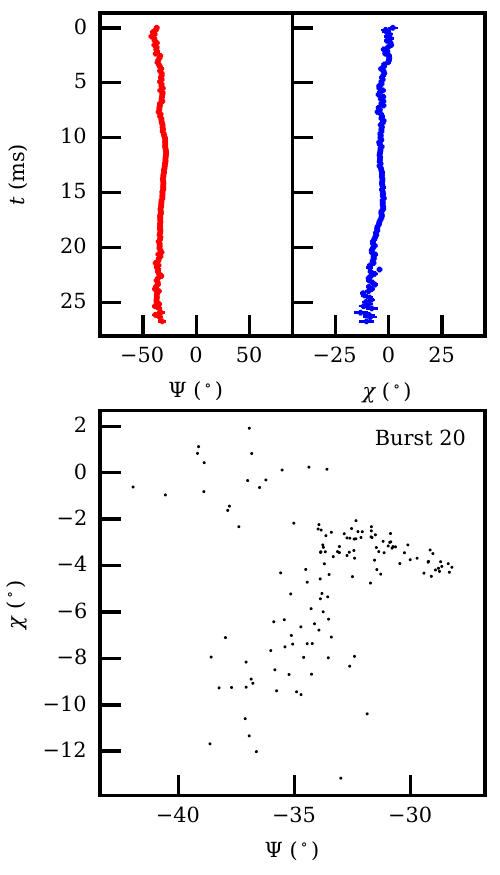}
\includegraphics[width=0.19\linewidth]{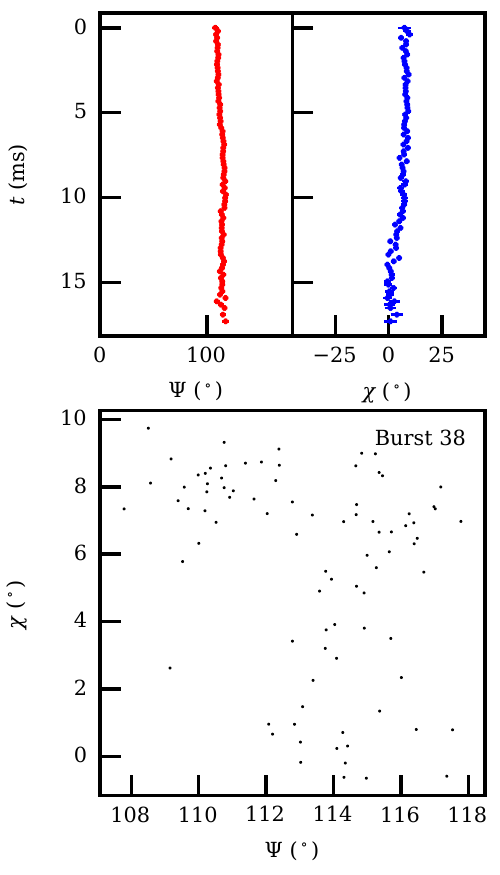}
\includegraphics[width=0.19\linewidth]{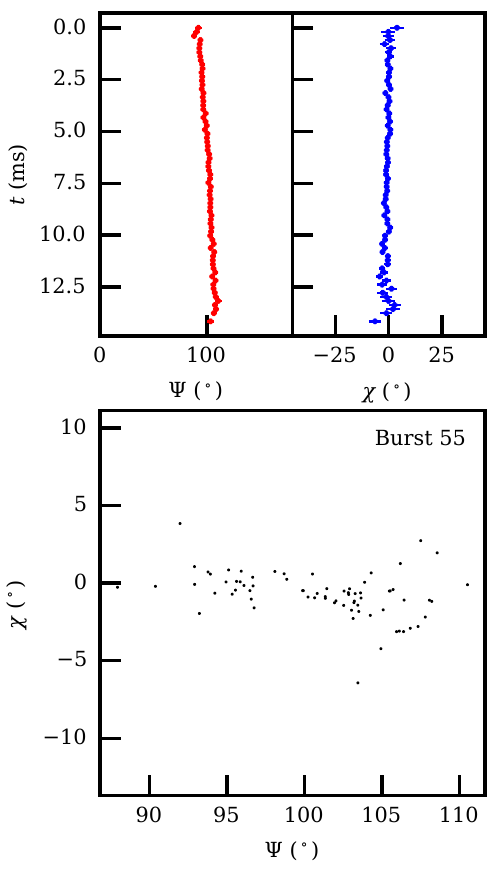}
\includegraphics[width=0.19\linewidth]{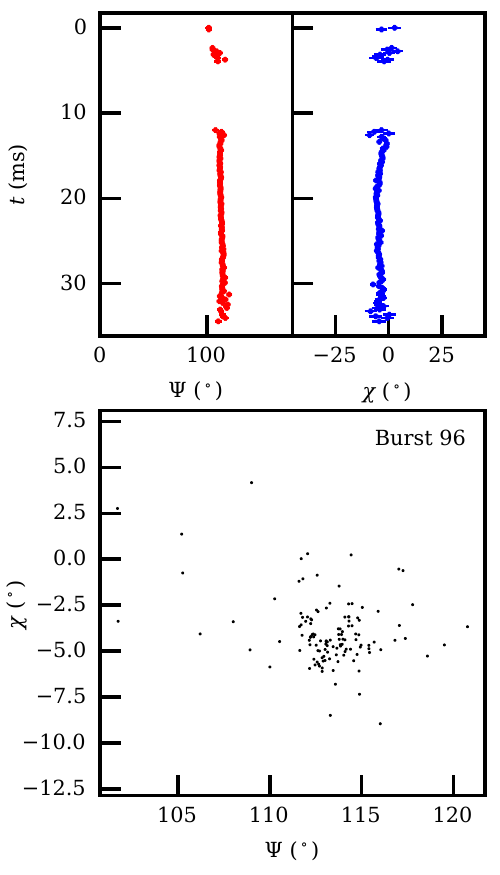}\\
\includegraphics[width=0.19\linewidth]{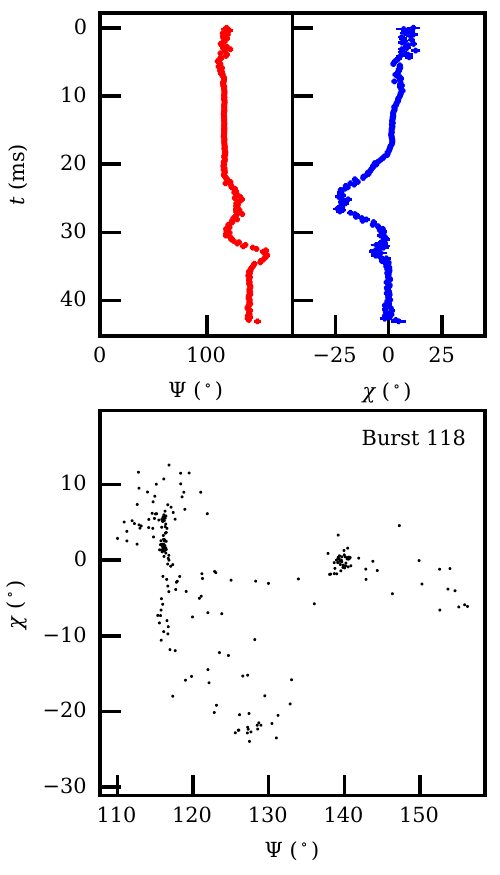}
\includegraphics[width=0.19\linewidth]{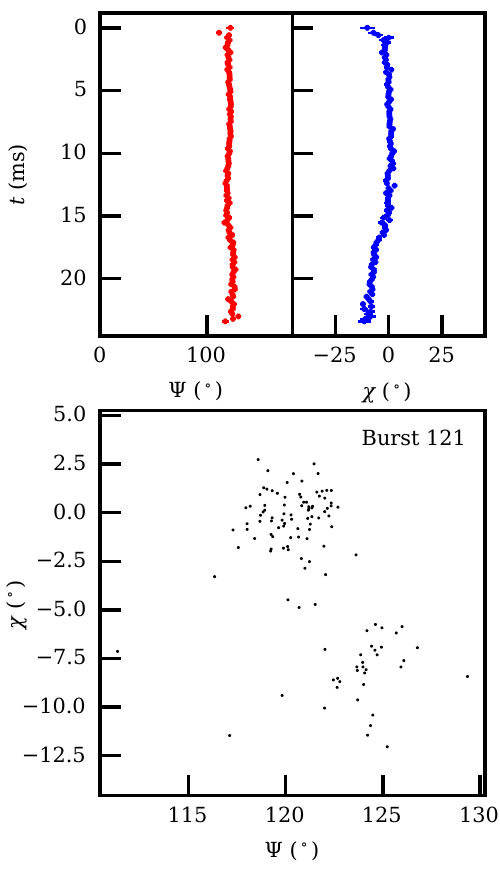}
\includegraphics[width=0.19\linewidth]{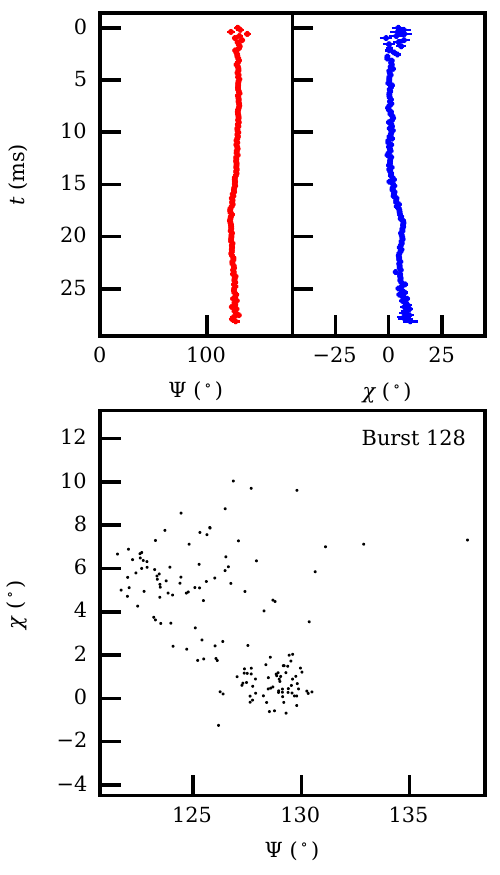}
\includegraphics[width=0.19\linewidth]{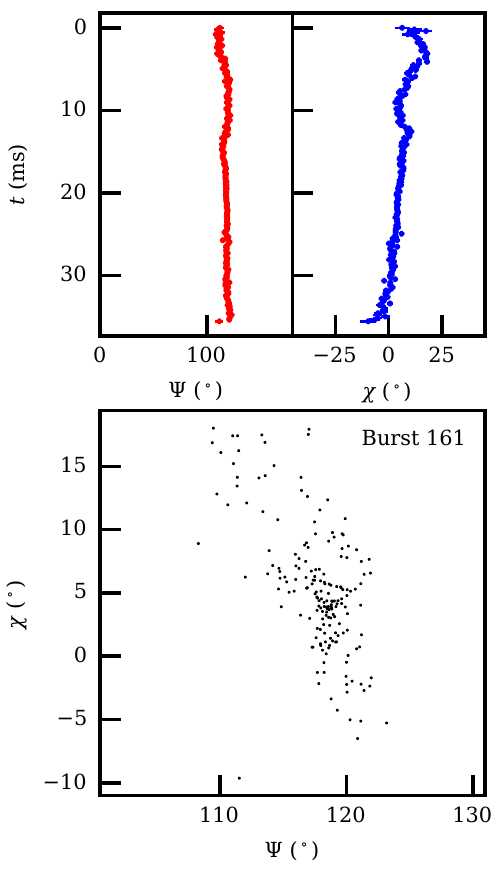}
\includegraphics[width=0.19\linewidth]{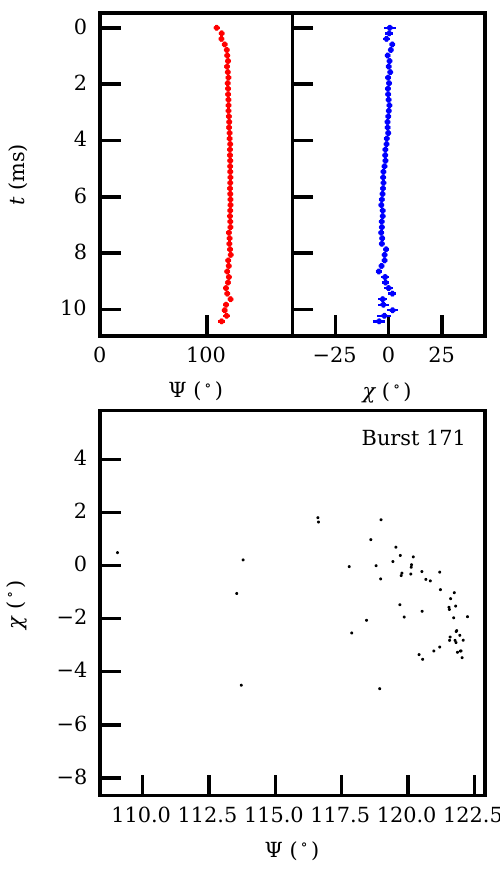}\\
\includegraphics[width=0.19\linewidth]{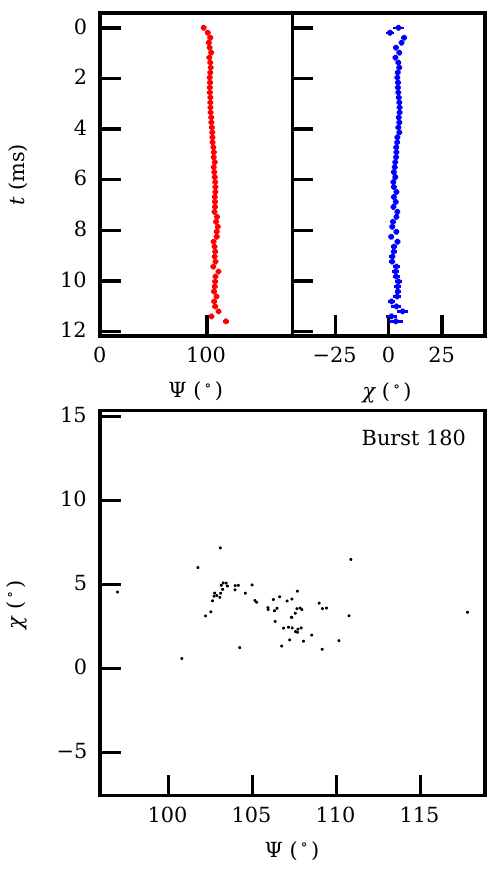}
\includegraphics[width=0.19\linewidth]{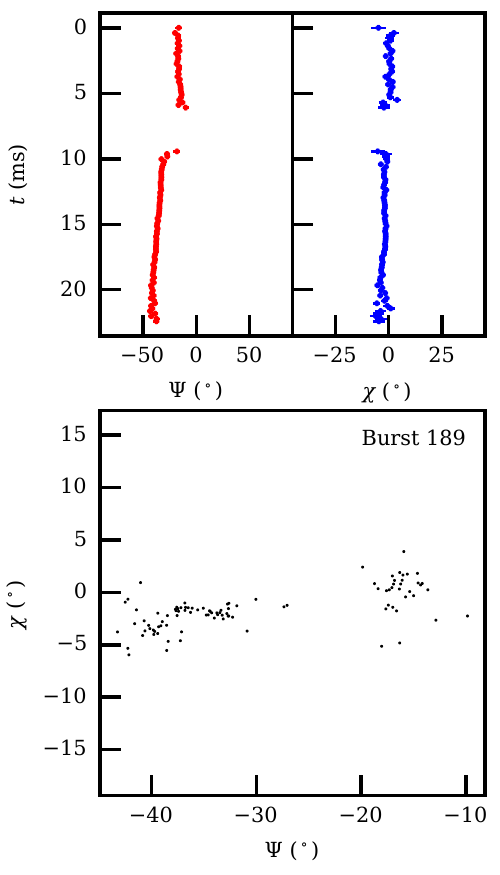}
\includegraphics[width=0.19\linewidth]{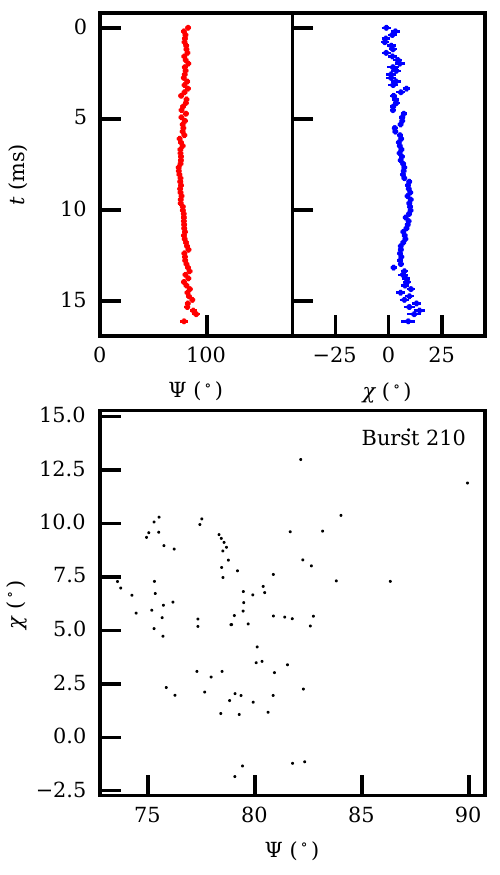}
\includegraphics[width=0.19\linewidth]{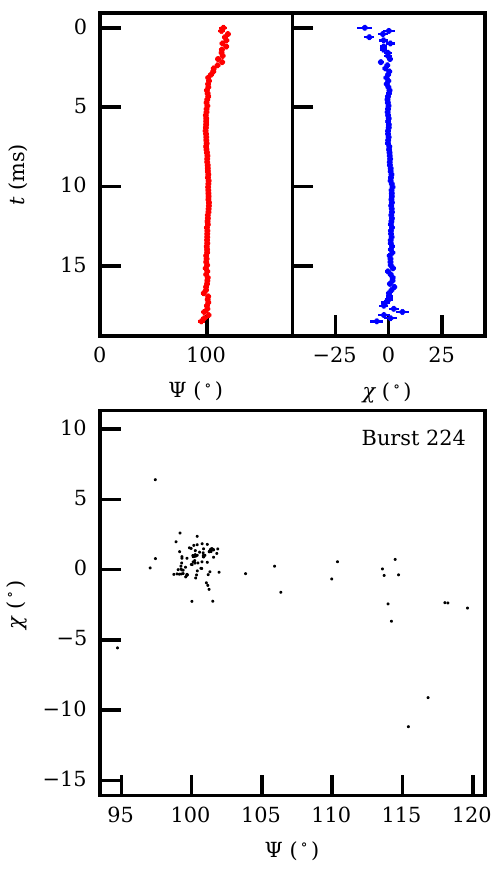}
\includegraphics[width=0.19\linewidth]{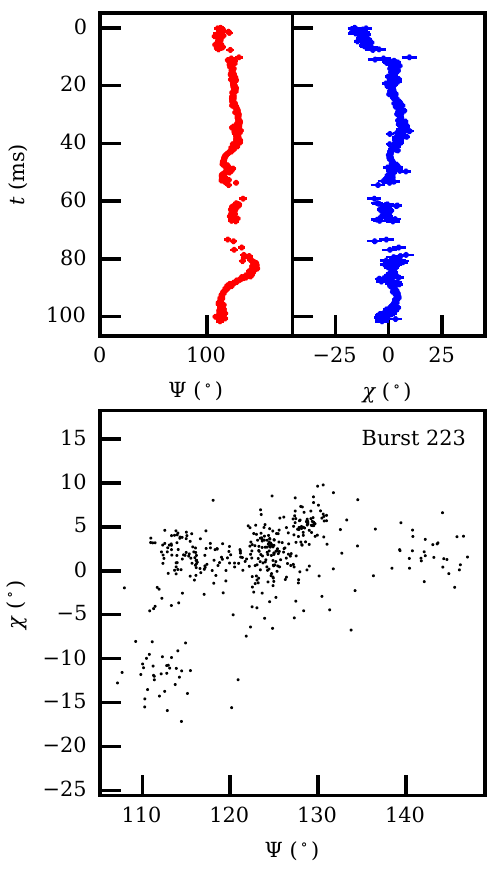}\\
\includegraphics[width=0.19\linewidth]{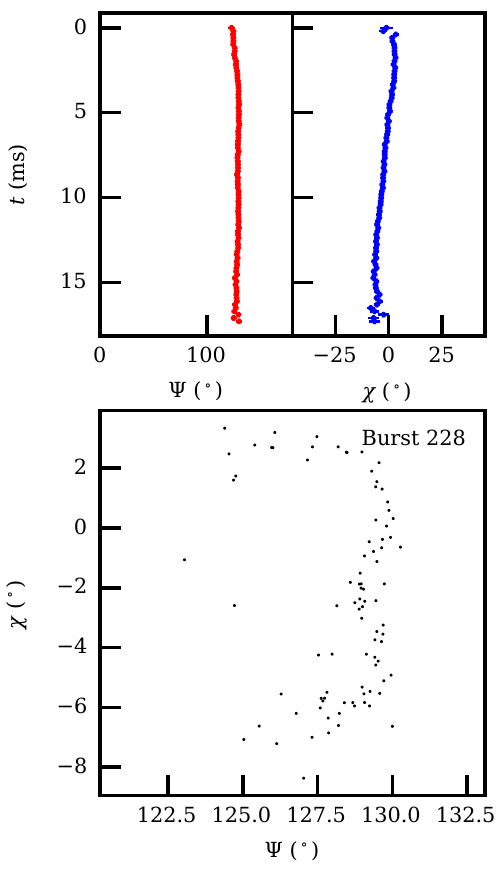}
\includegraphics[width=0.19\linewidth]{poincare/FRB20201124A_tracking-M01_0402_59484.8431332383_1982_2290_poincare.pdf}
\includegraphics[width=0.19\linewidth]{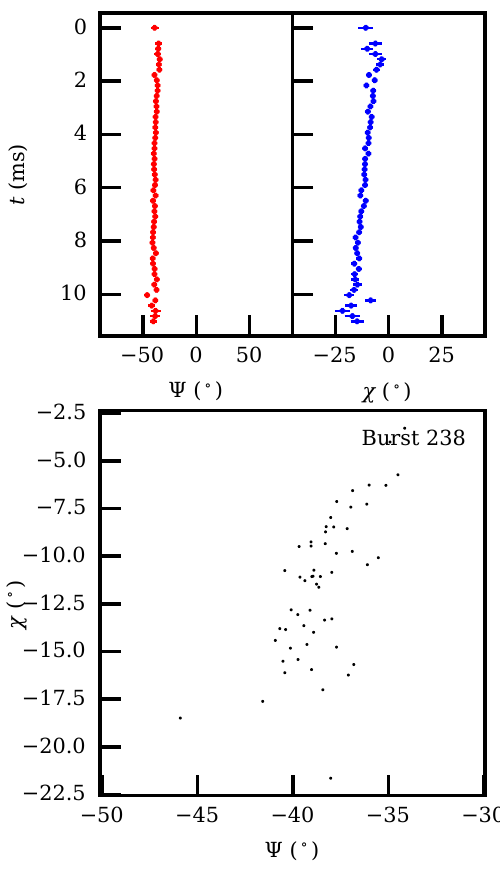}
\includegraphics[width=0.19\linewidth]{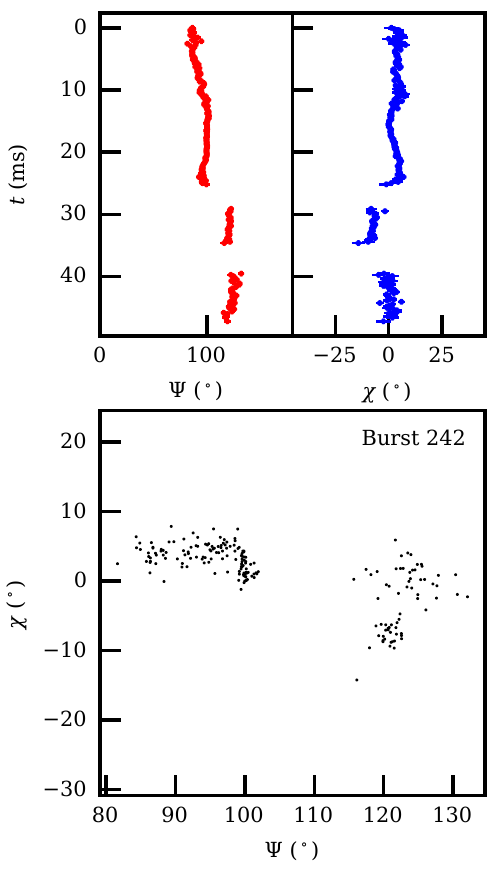}
\includegraphics[width=0.19\linewidth]{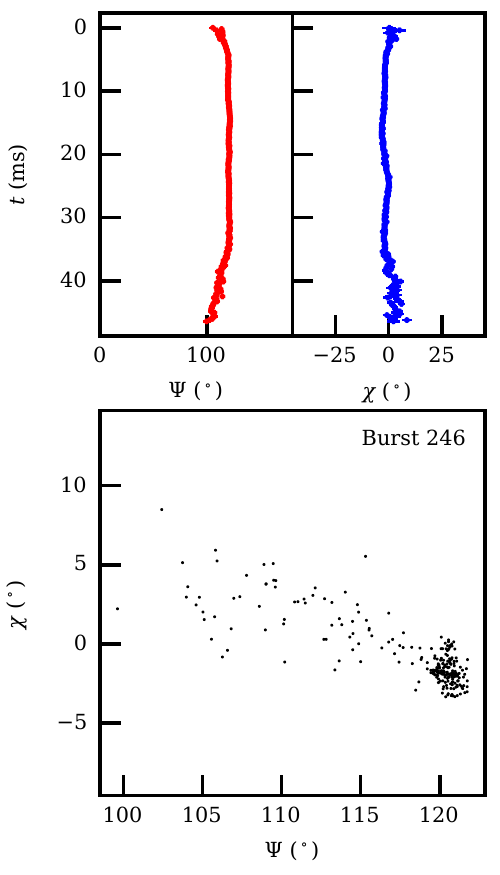}
\caption{$\Psi$ and $\chi$ curves of all bursts with $\Delta\Psi >20^\circ$, $\Delta\chi >10^\circ$ and $\mathrm{S/N}>200$. The denotation is same as Fig.~\ref{fig:poincare_example}.}
\label{fig:poincare_all}
\end{figure*}
\addtocounter{figure}{-1}
\begin{figure*}
\centering
\includegraphics[width=0.19\linewidth]{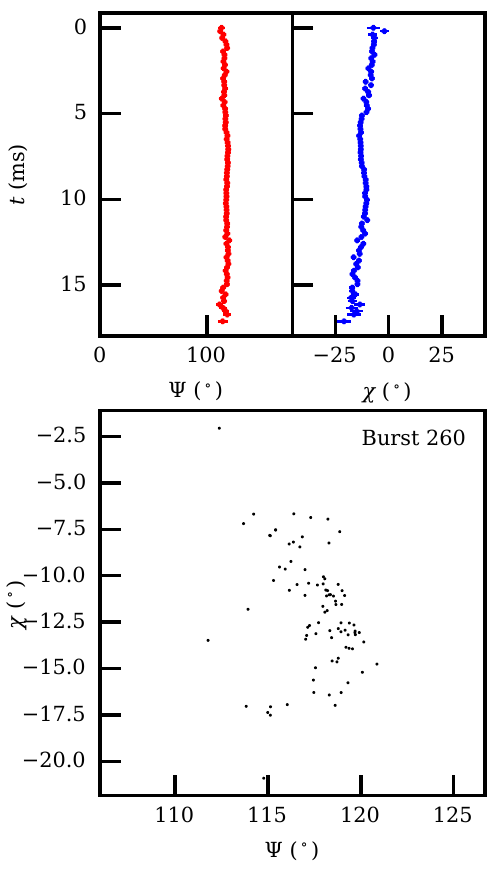}
\includegraphics[width=0.19\linewidth]{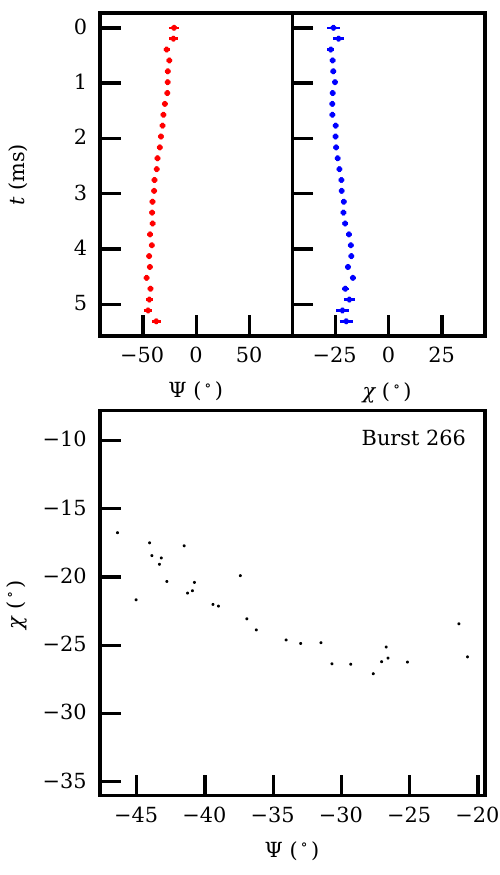}
\includegraphics[width=0.19\linewidth]{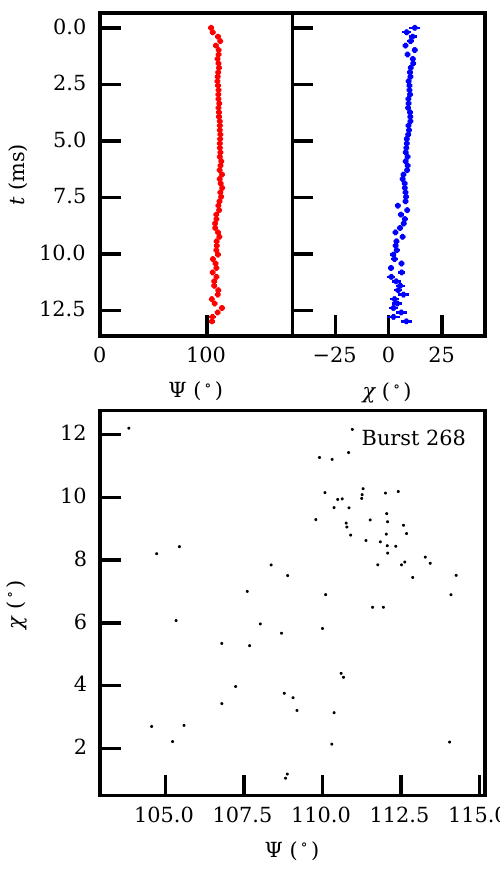}
\includegraphics[width=0.19\linewidth]{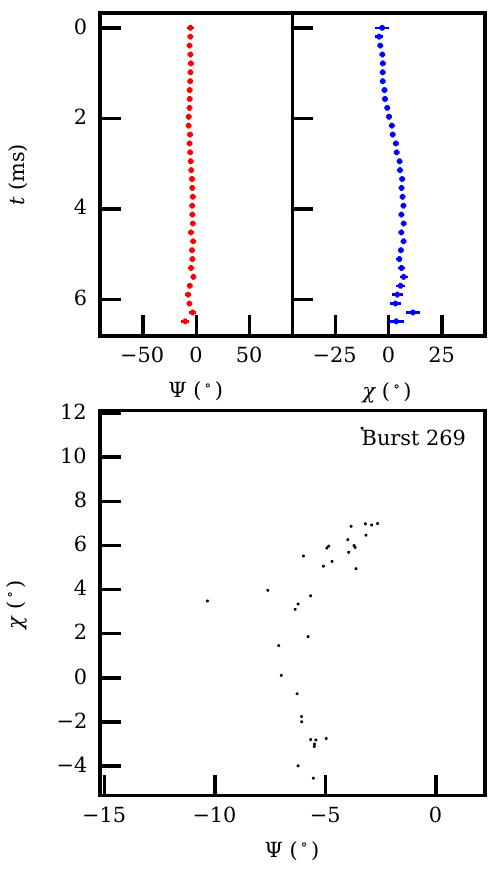}
\includegraphics[width=0.19\linewidth]{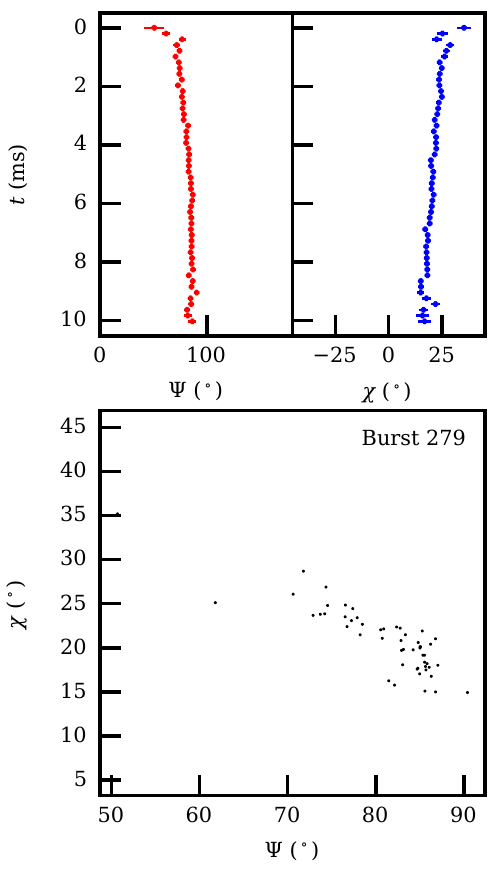}\\
\includegraphics[width=0.19\linewidth]{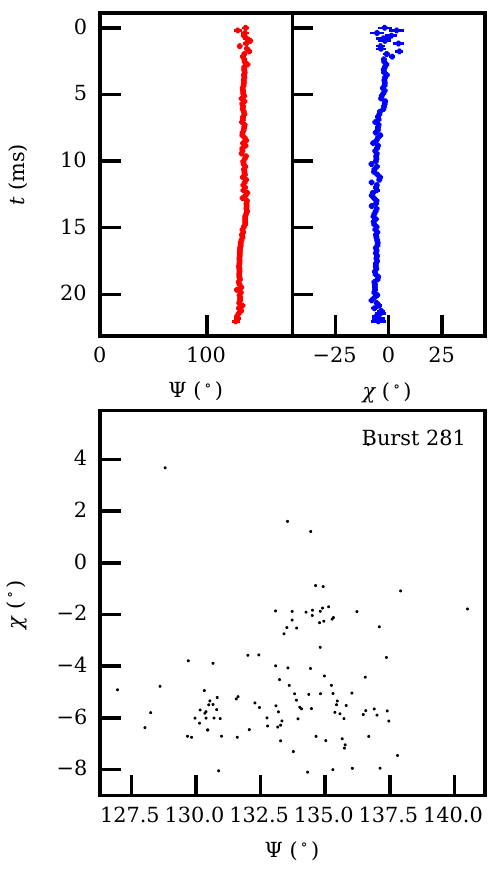}
\includegraphics[width=0.19\linewidth]{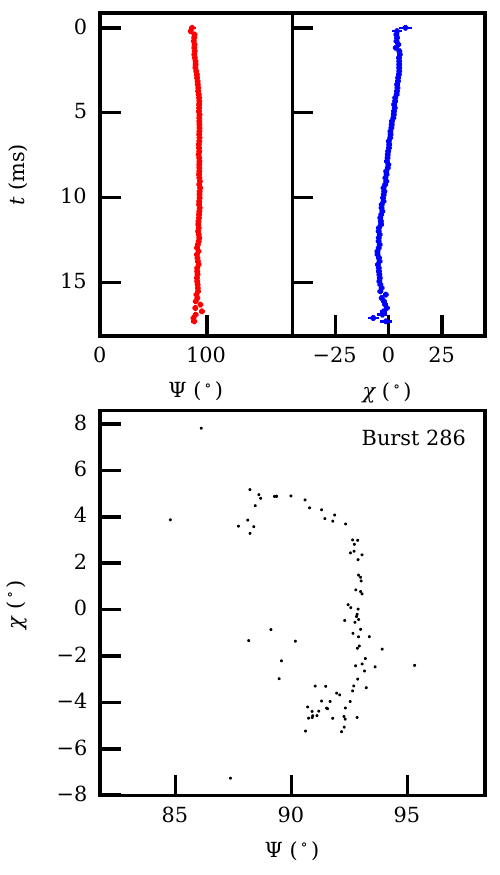}
\includegraphics[width=0.19\linewidth]{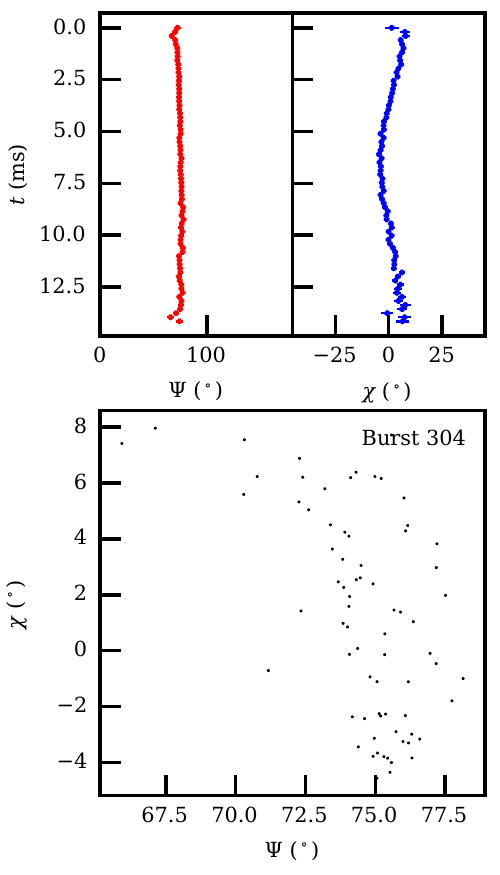}
\includegraphics[width=0.19\linewidth]{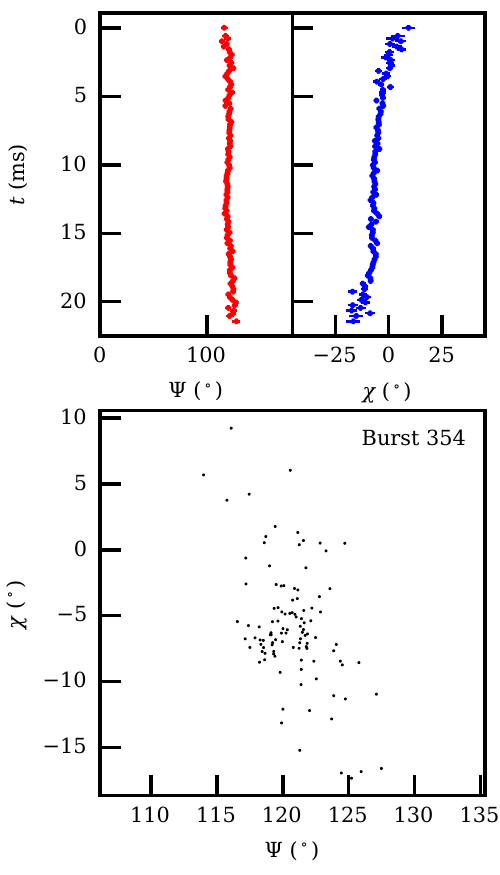}
\includegraphics[width=0.19\linewidth]{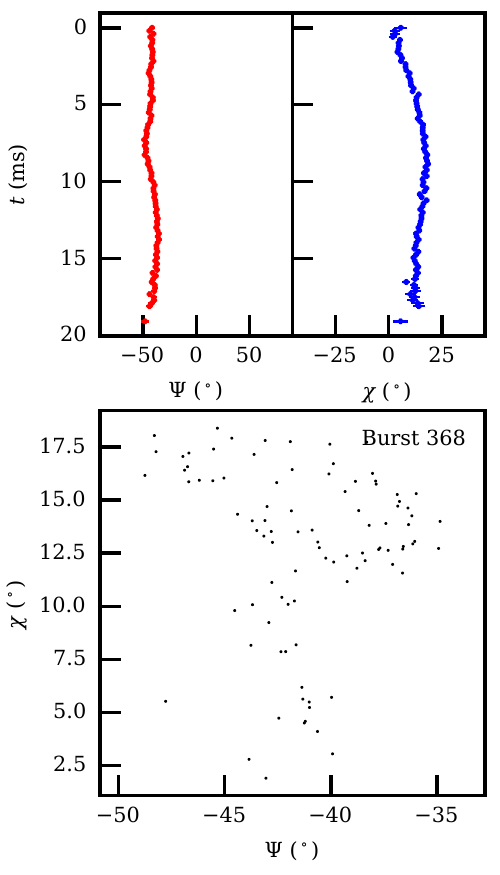}\\
\includegraphics[width=0.19\linewidth]{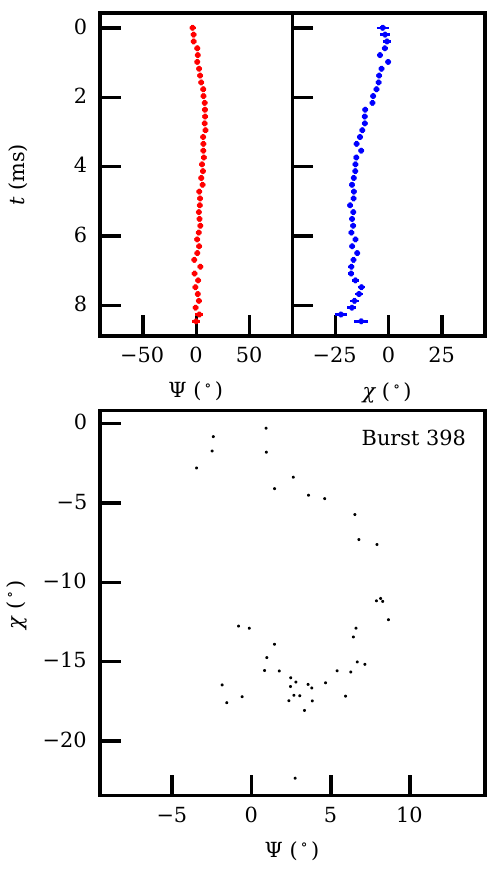}
\includegraphics[width=0.19\linewidth]{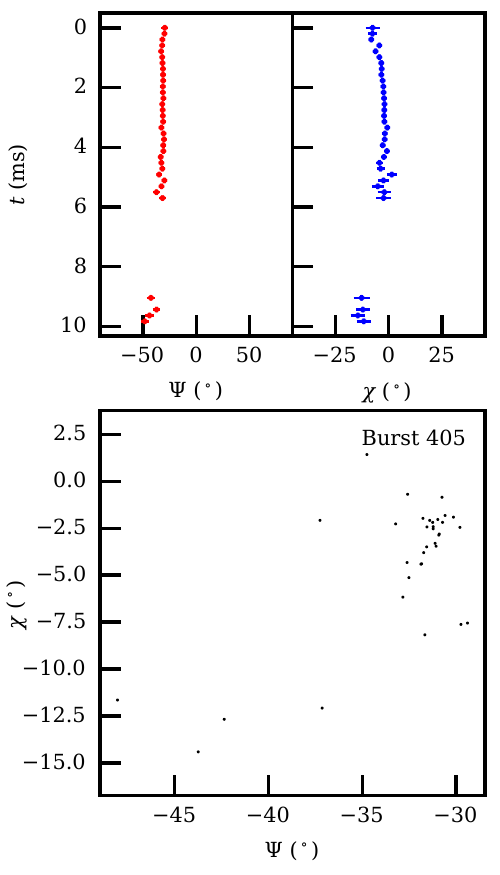}
\includegraphics[width=0.19\linewidth]{poincare/FRB20201124A_tracking-M01_0178_59485.7951478652_1432_2327_poincare.pdf}
\includegraphics[width=0.19\linewidth]{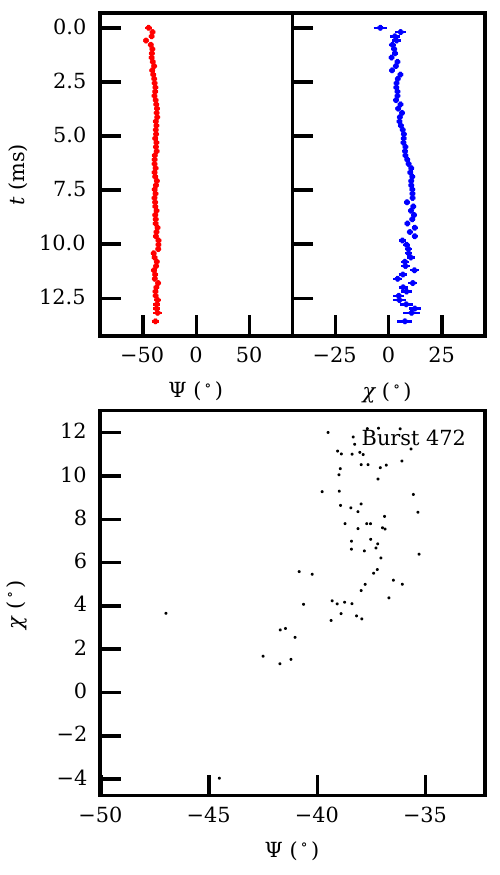}
\includegraphics[width=0.19\linewidth]{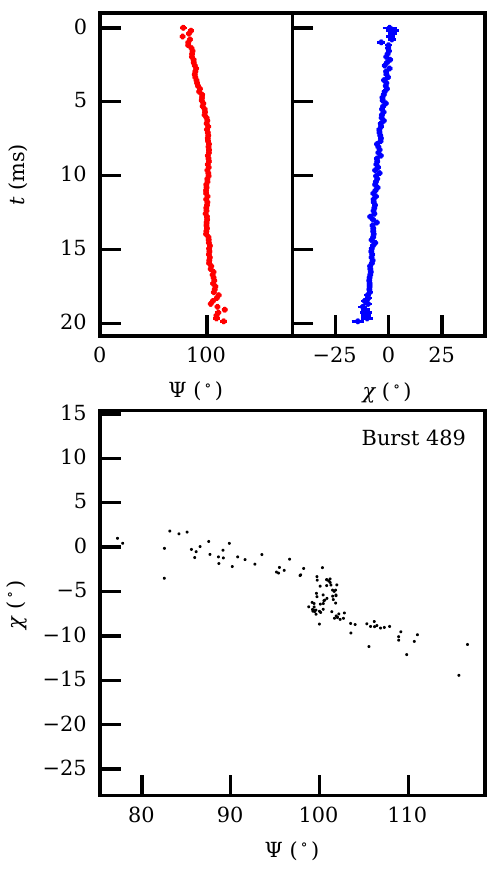}\\
\includegraphics[width=0.19\linewidth]{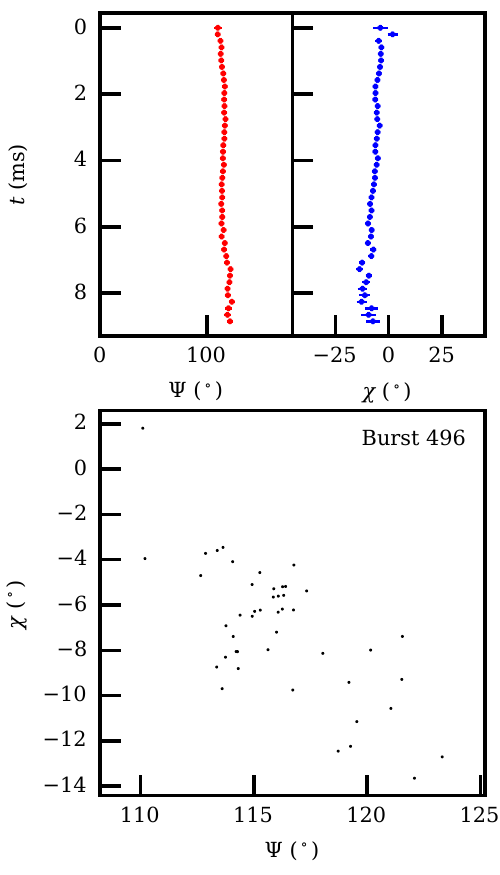}
\includegraphics[width=0.19\linewidth]{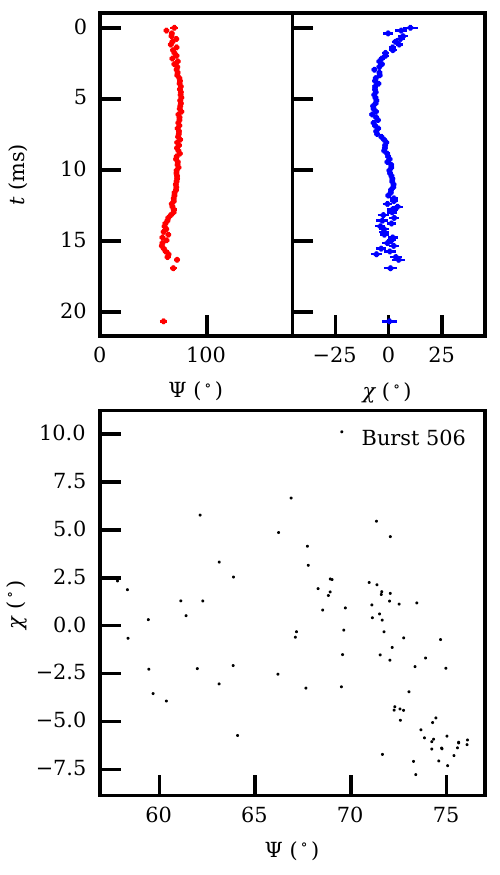}
\includegraphics[width=0.19\linewidth]{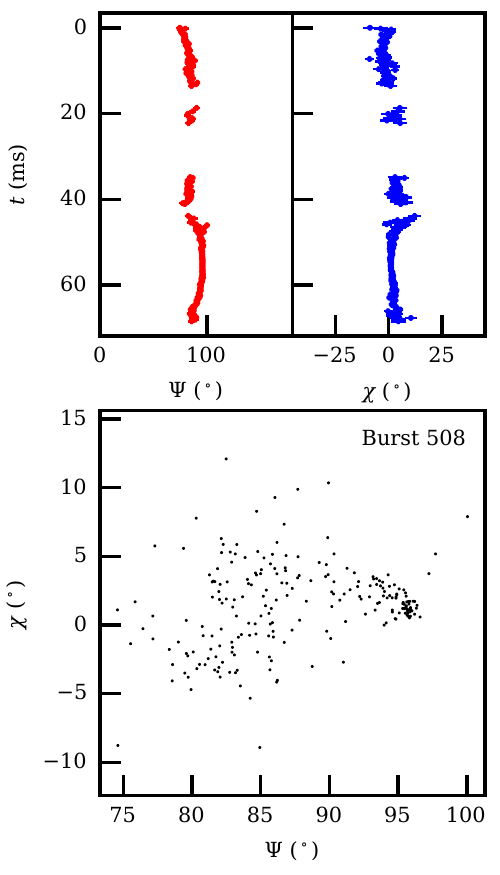}
\includegraphics[width=0.19\linewidth]{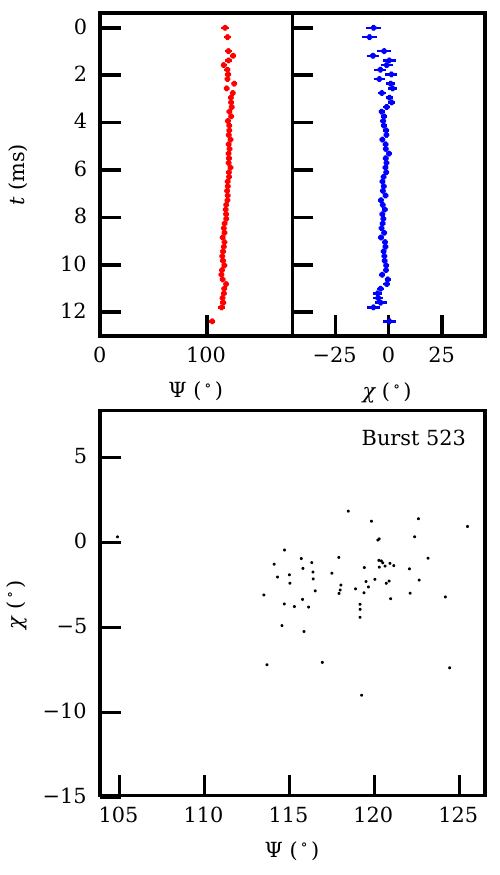}
\includegraphics[width=0.19\linewidth]{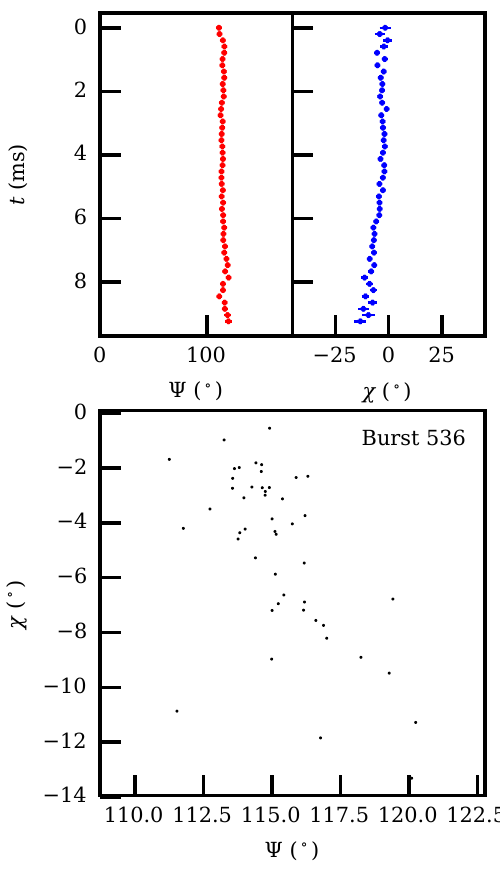}
\caption{\textit{continued.}}
\end{figure*}
\addtocounter{figure}{-1}
\begin{figure*}
\centering
\includegraphics[width=0.19\linewidth]{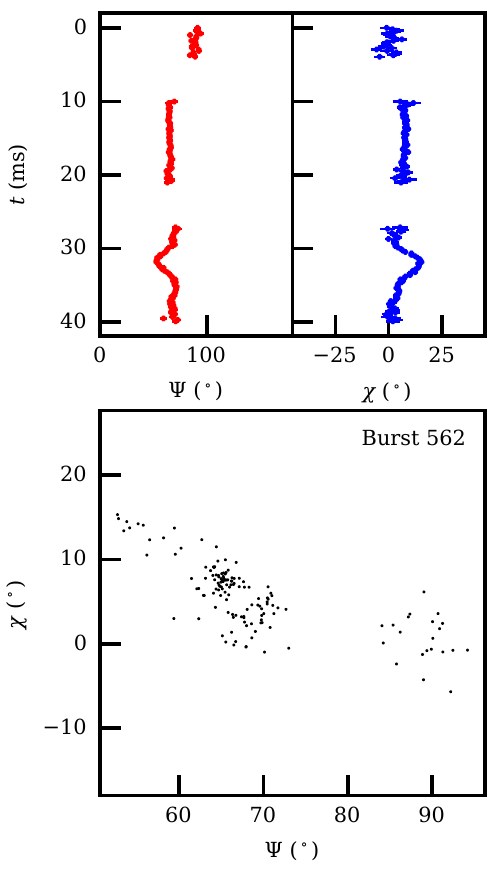}
\includegraphics[width=0.19\linewidth]{poincare/FRB20201124A_tracking-M01_0339_59485.8071716163_3036_3654_poincare.pdf}
\includegraphics[width=0.19\linewidth]{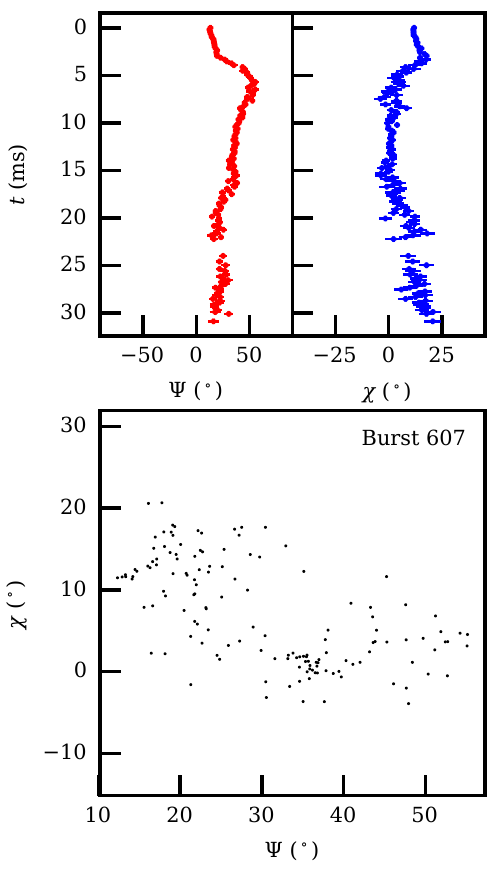}
\includegraphics[width=0.19\linewidth]{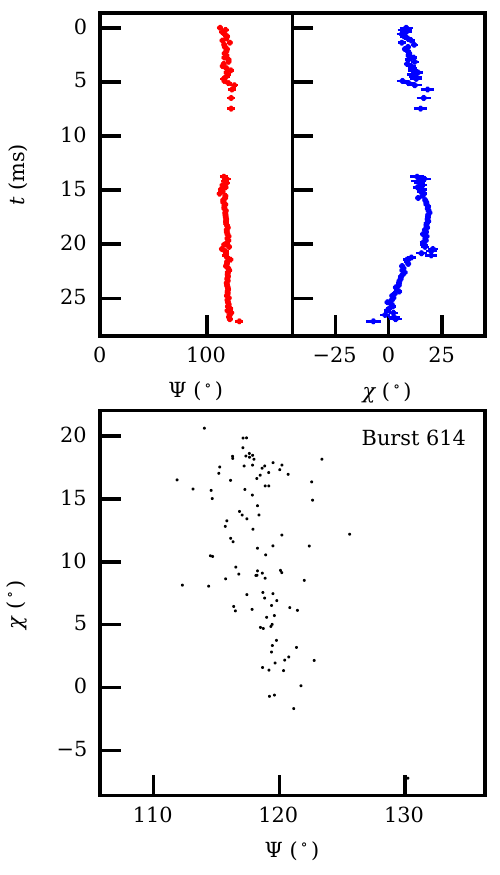}
\includegraphics[width=0.19\linewidth]{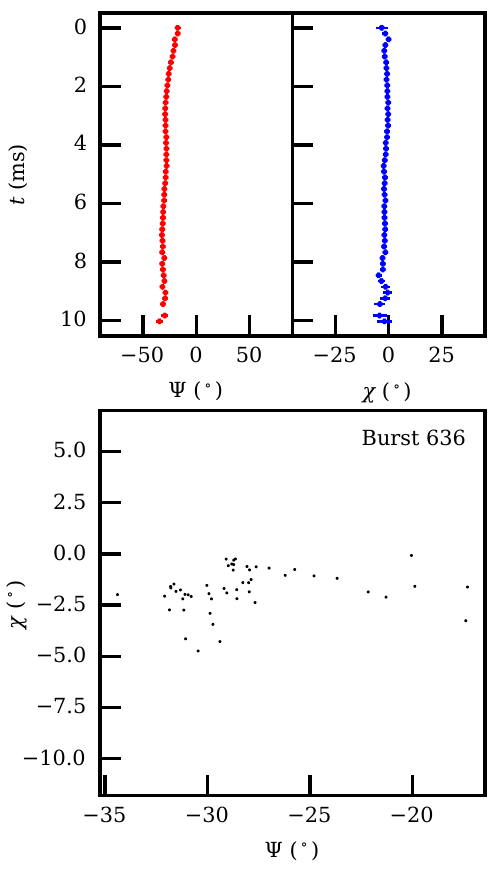}\\
\includegraphics[width=0.19\linewidth]{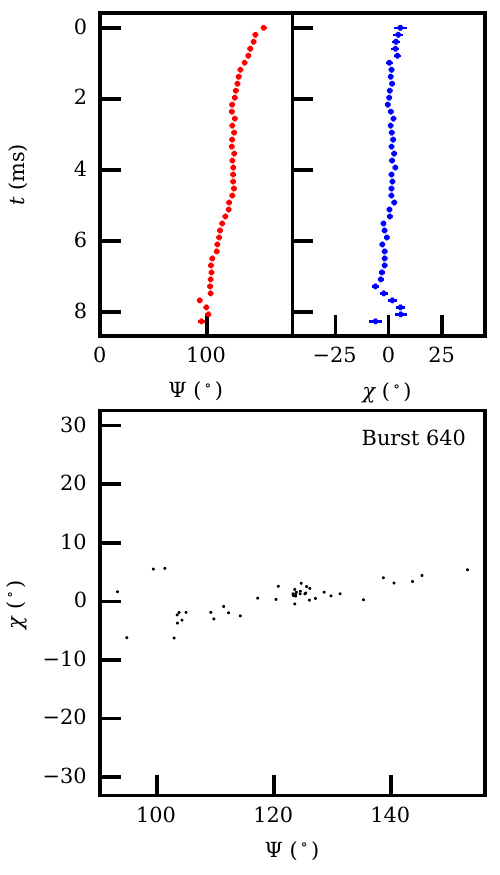}
\includegraphics[width=0.19\linewidth]{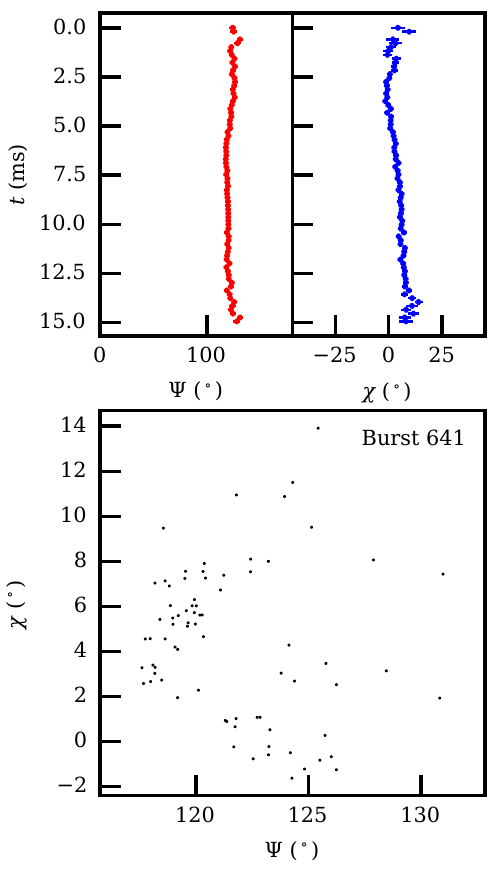}
\includegraphics[width=0.19\linewidth]{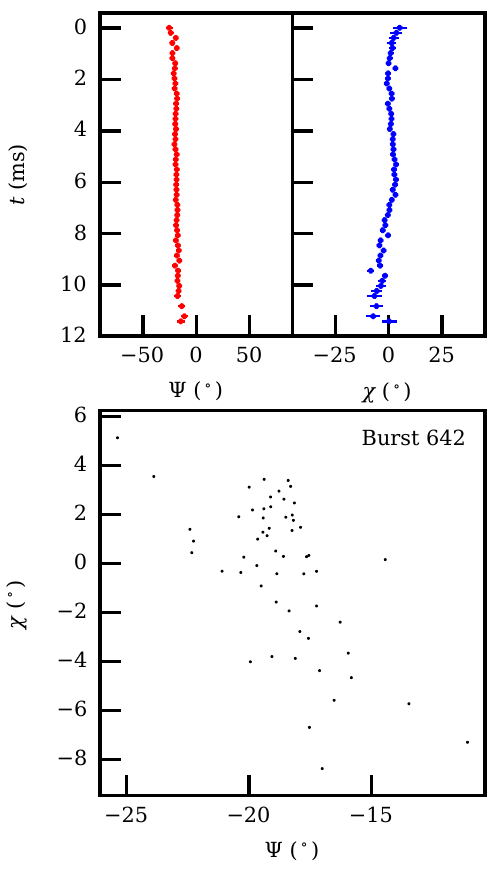}
\includegraphics[width=0.19\linewidth]{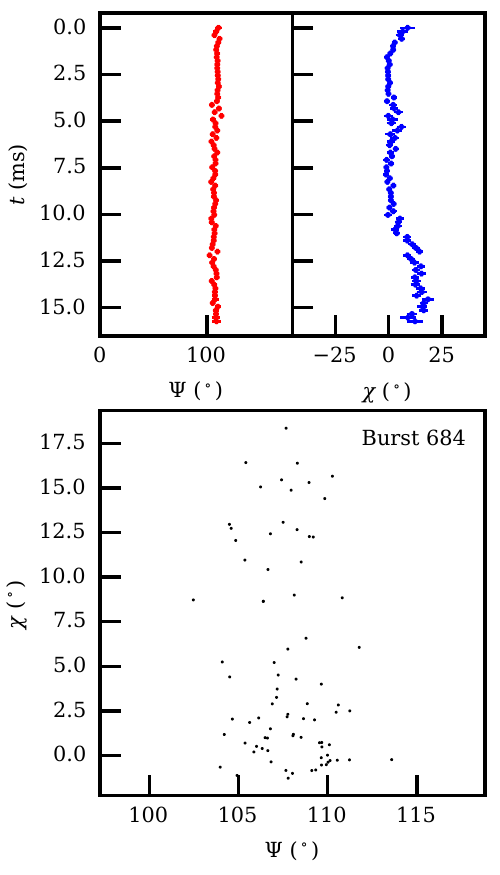}
\includegraphics[width=0.19\linewidth]{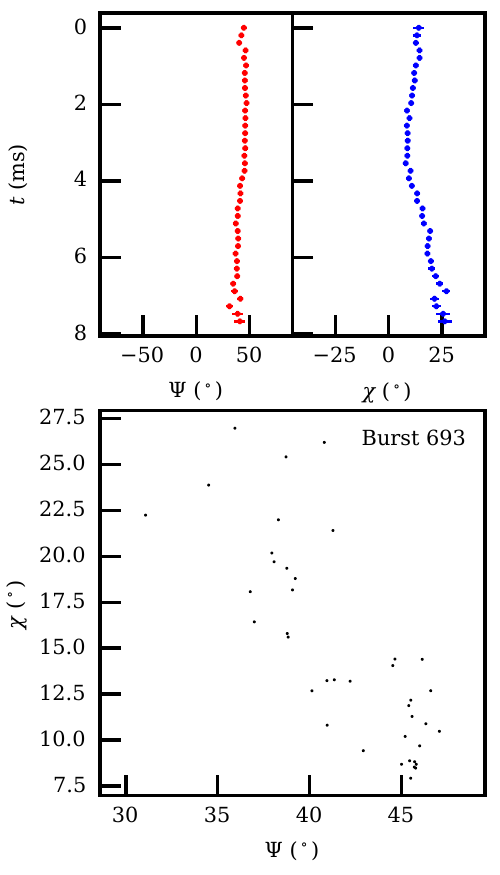}\\
\includegraphics[width=0.19\linewidth]{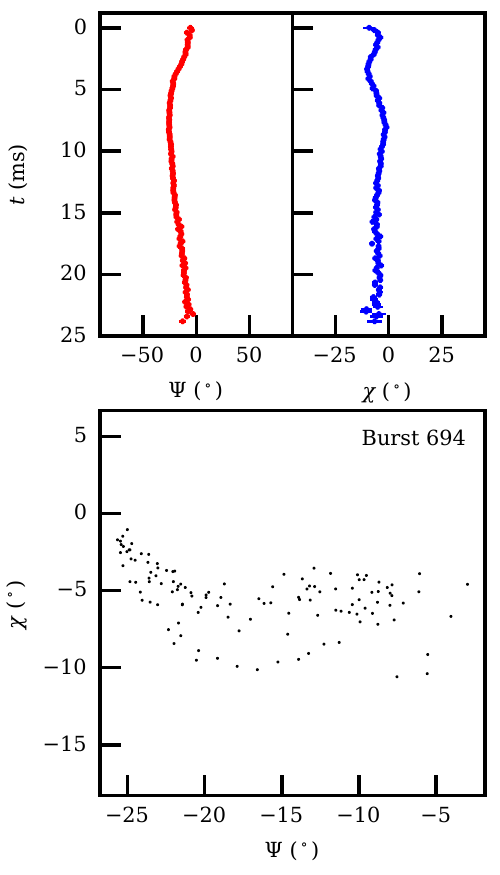}
\includegraphics[width=0.19\linewidth]{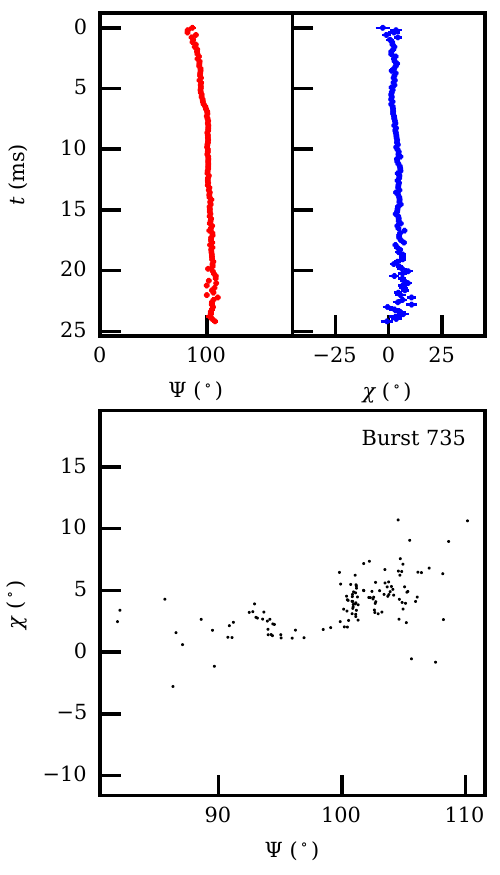}
\includegraphics[width=0.19\linewidth]{poincare/FRB20201124A_tracking-M01_0496_59485.8188957507_1881_2310_poincare.pdf}
\includegraphics[width=0.19\linewidth]{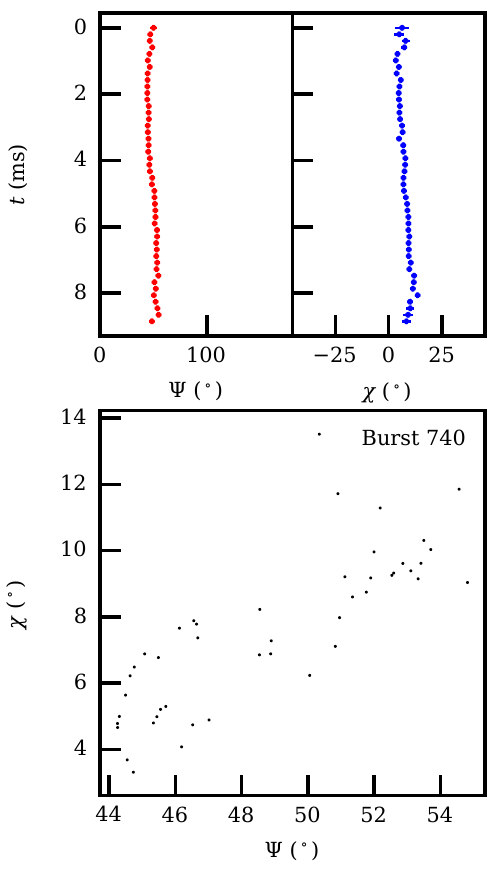}
\includegraphics[width=0.19\linewidth]{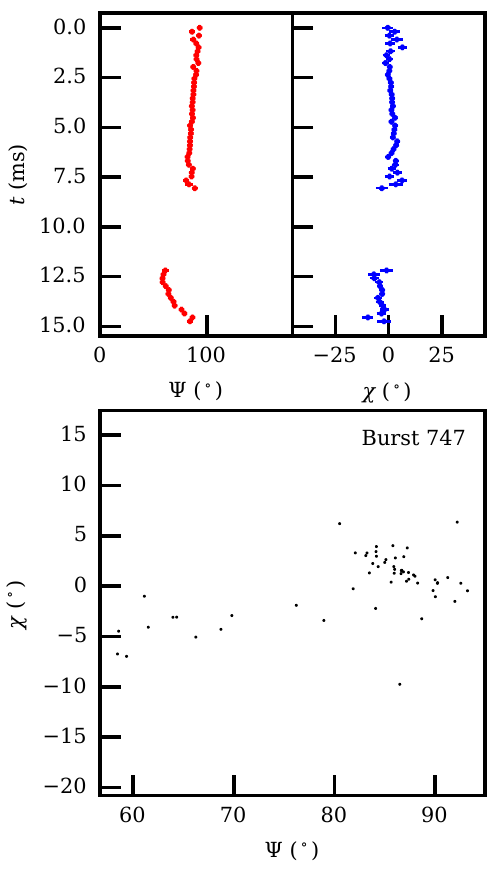}\\
\includegraphics[width=0.19\linewidth]{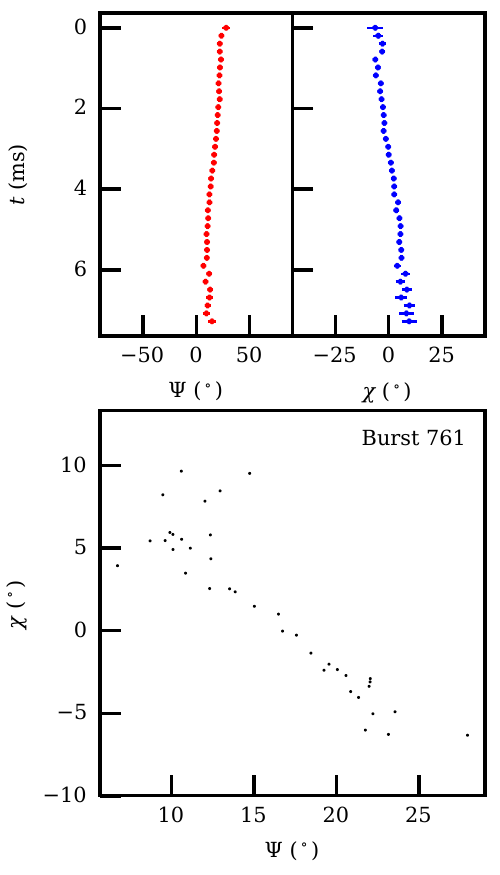}
\includegraphics[width=0.19\linewidth]{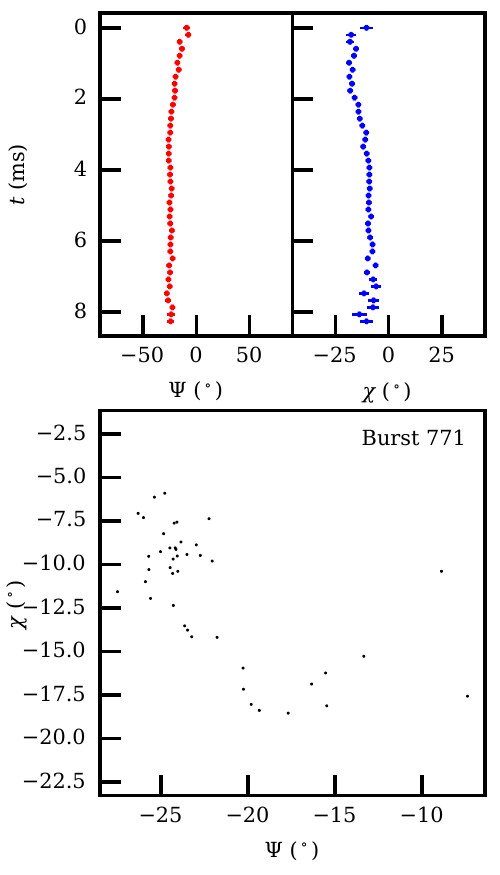}
\includegraphics[width=0.19\linewidth]{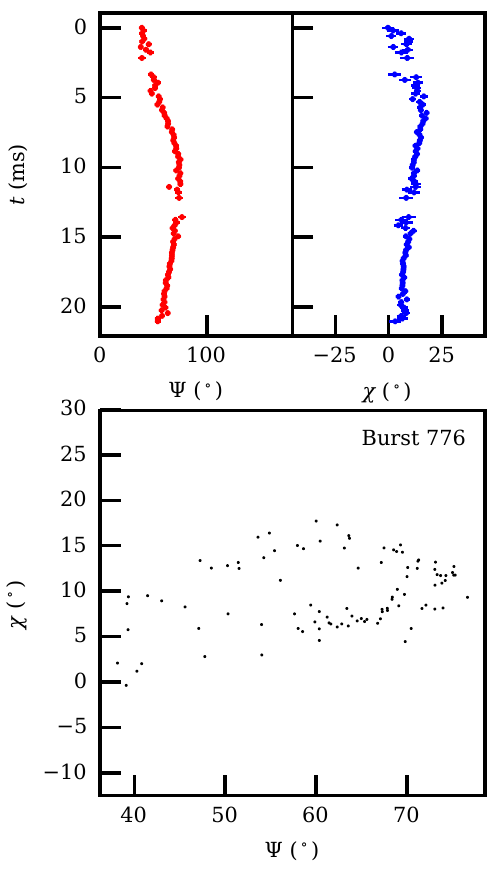}
\includegraphics[width=0.19\linewidth]{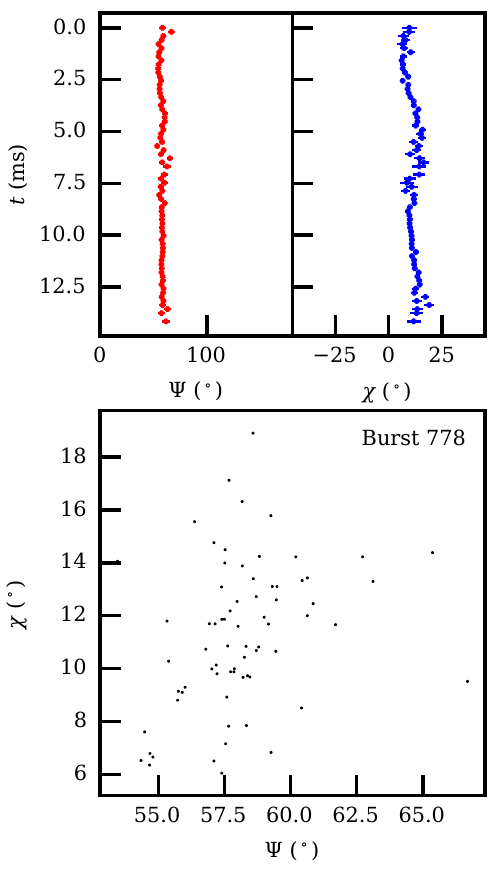}
\caption{\textit{continued.}}
\end{figure*}

\label{lastpage}

\end{document}